\documentclass[5p,preprint,floatfix,twocolumn]{elsarticle}

\usepackage{color}
\usepackage[para]{threeparttable} % for table footnotes
\usepackage{amsmath}
\usepackage{graphicx}
\usepackage{booktabs}
\usepackage{subfig}
\usepackage{miller}
\usepackage{placeins}

\usepackage[utf8]{inputenc}
\usepackage{hyperref}

\biboptions{sort&compress}

\makeatletter
\def\ps@pprintTitle{%
      \let\@oddhead\@empty
       \let\@evenhead\@empty
       \def\@oddfoot{\reset@font\hfil\thepage\hfil}
       \let\@evenfoot\@oddfoot
    }
\makeatother

\begin{document}

\begin{frontmatter}
\title{Dynamical stability of radiation-induced C15 clusters in iron}

\author[hu]{J. Byggmästar\corref{cor1}}
\ead{jesper.byggmastar@helsinki.fi}
\author[hu]{F. Granberg}
\cortext[cor1]{Corresponding author}
\address[hu]{Department of Physics, P.O. Box 43, FI-00014 University of Helsinki, Finland}

\begin{abstract}
    Density functional theory predicts clusters in the form of the C15 Laves phase to be the most stable cluster of self-interstitials in iron at small sizes. The C15 clusters can form as a result of irradiation, but their prevalence and survival in harsh irradiation conditions have not been thoroughly studied. Using a new bond-order potential optimised for molecular dynamics simulations of radiation damage, we explore the dynamical stability of the C15 clusters in iron under irradiation conditions. We find that small C15 clusters make up 5--20\% of the interstitial clusters formed directly in cascades. In continuous irradiation, C15 clusters are frequently formed, after which they remain highly stable and grow by absorbing nearby single interstitial atoms. Growth of C15 clusters ultimately leads to collapse into dislocation loops, most frequently into $1/2\hkl<111>$ loops and only rarely collapsing into \hkl<100> loops at low temperatures. The population, size, and collapse of C15 clusters during continuous irradiation correlates well with their formation energies relative to dislocation loops calculated at zero Kelvin.
\end{abstract}


\end{frontmatter}

\section{Introduction}
\label{sec:intro}

Irradiation of iron and iron-based alloys results in accumulation of defects and defect clusters. Self-interstitial atoms (SIAs) and vacancies cluster together as dislocation loops, voids, and other defect clusters. Unique to iron, small SIA clusters are most stable in the form of agglomerates of C15 Laves crystals, coherently embedded in the bcc structure as predicted by density functional theory~\cite{marinica_irradiation-induced_2012}. Molecular dynamics (MD) simulations show that C15 clusters can form in radiation-induced collision cascades~\cite{marinica_irradiation-induced_2012,byggmastar_effects_2018} or by direct clustering of migrating SIAs~\cite{chartier_rearrangement_2019}, similar to the conditions during electron irradiation. The presence of these highly stable and stationary three-dimensional clusters sets iron apart from non-magnetic bcc metals, like tungsten, where rapidly migrating $1/2\hkl<111>$ loops are the dominant radiation-induced clusters~\cite{yi_situ_2013,yi_characterisation_2015,sand_high-energy_2013,byggmastar_collision_2019}. At larger cluster sizes, however, dislocation loops eventually become the most stable configuration in iron too~\cite{alexander_ab_2016}. The small size of the C15 clusters make them difficult to observe experimentally, and the first experimental measurements are yet to be published. Hence, the presence and effects of C15 clusters on the microstructural evolution of iron under irradiation still remains unclear.

Atomistic simulations can be used to provide useful predictions of the formation, stability, and evolution of defect clusters, such as the C15 clusters. Even though the relative stability at zero temperature of various defect clusters can be estimated from density functional theory, observing the formation and evolution of C15 clusters requires larger-scale classical MD simulations. One critical restriction is, however, the lack of interatomic potentials that correctly predict the C15 clusters as the most stable interstitial-rich clusters in iron. To date, only the embedded atom method potentials by Marinica et al.~\cite{marinica_irradiation-induced_2012,malerba_comparison_2010} correctly stabilise the C15 cluster in relation to parallel $\hkl<111>$ interstitial configurations. Nevertheless, these potentials also incorrectly predicts $\hkl<100>$ dislocation loops to become more stable than $1/2\hkl<111>$ loops at sizes larger than around 80 SIAs~\cite{malerba_comparison_2010}. This makes it difficult to assess the reliability of simulated cluster transformations, for example transformations from C15 to dislocation loops due to cascade overlap~\cite{byggmastar_collision_2019}, or collapse of growing C15 clusters into a dislocation loop of either type~\cite{zhang_formation_2015,chartier_rearrangement_2019}.

The aim of this article is to study the formation, stability, growth, and collapse of radiation-induced C15 clusters in iron. To achieve this, we first develop a new Tersoff-type analytical bond-order potential that, unlike all previous potentials for radiation damage, predicts the correct relative stability between C15 clusters and $1/2\hkl<111>$ and $\hkl<100>$ dislocation loops across all sizes. We use the new potential to investigate C15 clusters in iron and discuss our results in relation to previous observations and predictions by other interatomic potentials.

\section{Methods}
\label{sec:methods}

\subsection{Fitting the interatomic potential}

Our starting point when developing the potential was the existing analytical bond-order potential (ABOP) by M\"uller et al.~\cite{muller_analytic_2007} with the short-range addition by Björkas and Nordlund~\cite{bjorkas_comparative_2007}. The goal was to adjust the parameters in order to improve the energetics of single interstitials and vacancies, including migration energies, and to correctly stabilise small C15 clusters relative to dislocation loops in accordance with DFT predictions~\cite{marinica_irradiation-induced_2012,alexander_ab_2016}. To reproduce the stability of C15 clusters, we found it helpful to monitor the difference in cohesive energy and lattice mismatch between bulk C15 and bcc iron~\cite{dezerald_stability_2014}. The short-range connection to the ZBL potential was also optimised to better reproduce the many-body repulsion at intermediate interatomic distances. A more detailed description of the fitting strategy along with extensive benchmarking of the optimised ABOP is provided in the Supplementary material available online.

\subsection{Molecular statics and dynamics simulations}

Static energy minimisations and phonon calculations were carried out using \textsc{lammps}~\cite{plimpton_fast_1995} within the Atomic Simulation Environment (ASE)~\cite{larsen_atomic_2017}. For molecular dynamics simulations, we used the \textsc{parcas} code~\cite{nordlund_molecular_1995,nordlund_defect_1998}.

The formation energies of SIA clusters were calculated by minimsing the positions and pressure of a bcc system with around 54\,000 atoms. The collision cascade simulations were carried under identical conditions as outlined in~\cite{byggmastar_effects_2018} (from which the simulation data obtained with the AM04~\cite{ackland_development_2004} and M07-B~\cite{marinica_irradiation-induced_2012,malerba_comparison_2010,byggmastar_effects_2018} embedded atom method potentials were taken and further analysed in this work). All cells were quenched to zero Kelvin and an overall zero pressure over a period of 5 ps, to remove thermal displacements before the detailed analysis of the defect clusters. Wigner-Seitz analysis was used to isolate the interstitials and vacancies. Following that, all C15-like clusters were automatically identified by analysing the bond angles and planes of all interstitial clusters (after isolating the interstitial dumbbells using the Wigner-Seitz analysis, C15 clusters contain connected dumbbells with 60- and 120-degree angles on $\hkl<111>$ planes). This method of identifying C15-like clusters was found to be sufficiently robust when analysing highly damaged systems~\cite{byggmastar_effects_2018} as well as large amounts of overlapping cascade data~\cite{byggmastar_collision_2019}.

In the C15 growth simulations, an initial C15 cluster containing 17 SIAs was coherently inserted in a bcc system of 128\,000 atoms and relaxed at zero pressure at 500 K. Dumbbell self-interstitials in a randomly sampled $\hkl<110>$ direction were then inserted at intervals of 200 ps. The SIAs were placed at a random lattice site roughly 5--10 Å from the C15-bcc interface. After a new interstitial was added, the system was allowed to freely evolve in the $NVE$ ensemble. The simulation approach is similar to the C15 growth simulations carried out in Ref.~\cite{zhang_formation_2015}. During the 200 ps, the interstitial was likely to migrate towards the C15 cluster and eventually attach itself to it, resulting in a slowly growing C15 cluster. The simulations continued until the system contained 150 SIAs (corresponding to roughly 25 ns), before which the C15 cluster in almost all cases had collapsed or partially transformed into a dislocation loop. A total of 30 simulations were carried out in the ABOP to gather statistics. In addition, 20 growth simulations were performed with the M07-B potential for comparison.

\section{Results and discussion}

\subsection{Validation of the new interatomic potential}

\begin{table}
    \centering
    \begin{tabular}{llll}
        \toprule
         & Exp.~\cite{haynes_crc_2015} & DFT~\cite{dezerald_stability_2014} & ABOP \\
        \midrule
        \textbf{bcc} \\
         $E_\mathrm{coh}$ & 4.31 & & 4.35  \\
         $a$ & 2.866 & 2.84 & 2.86 \\
         $B$ & 169 & 162 & 171 \\
         \midrule
         \textbf{C15} \\
         $\Delta E$ & & 0.15 & 0.14\\
         $a$ & & 6.64 & 6.64 \\
         $B$ & & 150 & 157 \\
        \bottomrule
    \end{tabular}
    \caption{Cohesive energy (eV/atom), lattice constant (Å), and bulk modulus (GPa) of bulk bcc and C15 iron given by the ABOP and compared with experimental and DFT results.}
    \label{tab:C15bulk}
\end{table}

Before using the new interatomic potential to study the formation and growth of C15 clusters, we here shortly demonstrate its applicability by calculating properties relevant for simulations of radiation-induced defect clusters. In addition to the results presented below, the ABOP describes well the elastic and vibrational properties of bcc iron and the threshold displacement energies, making it suitable for collision cascade simulations. These and additional validation results are presented in the Supplementary material. We compare the results with two embedded atom method (EAM) potentials (AM04~\cite{ackland_development_2004} and M07-B~\cite{marinica_irradiation-induced_2012,malerba_comparison_2010,byggmastar_effects_2018}).

\begin{figure}
 \centering
 \includegraphics[width=\linewidth]{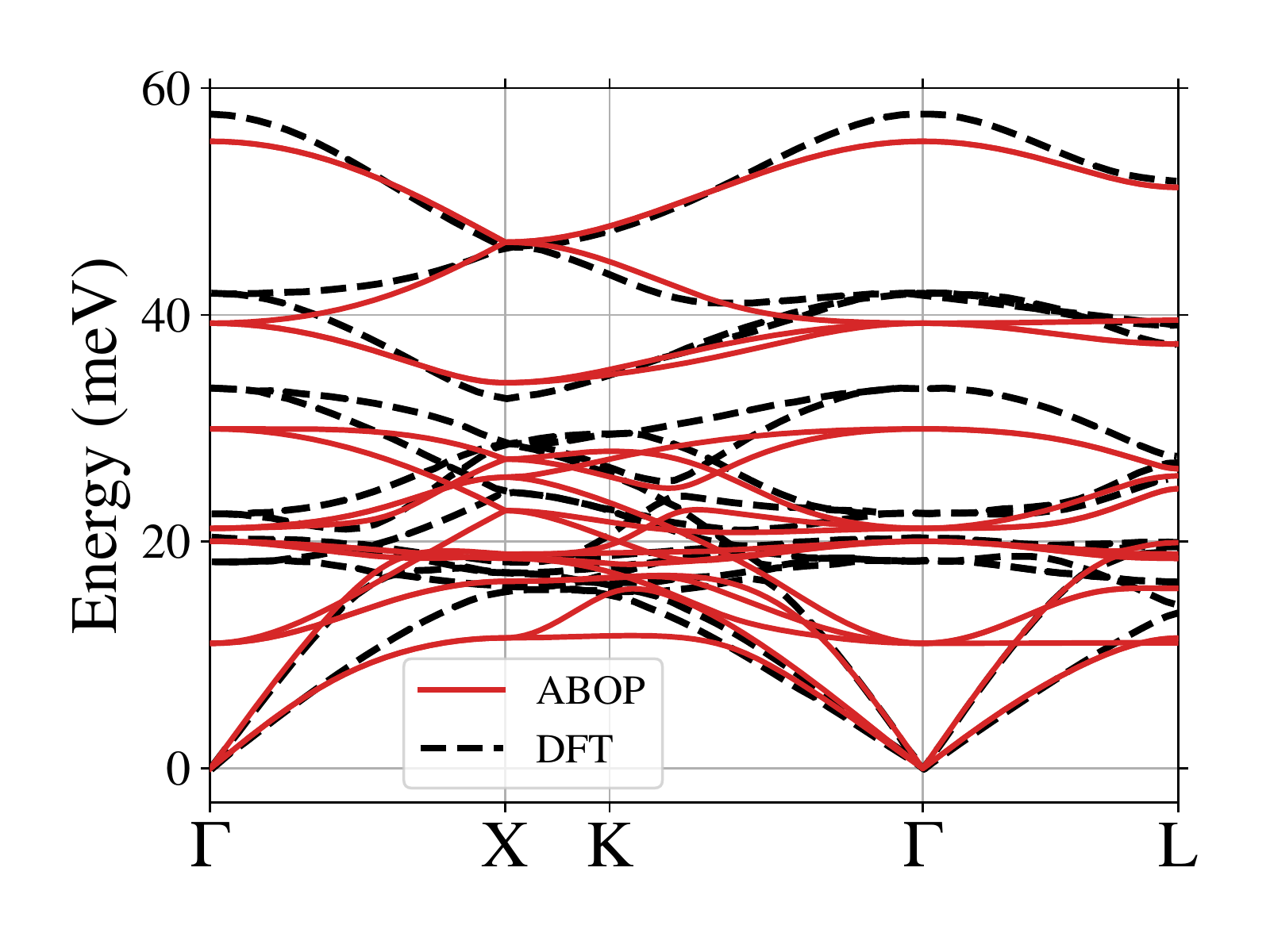}
 \caption{Phonon dispersion of bulk C15 Fe in the ABOP, compared with DFT results from Ref.~\cite{dezerald_stability_2014}.}
 \label{fig:phonon_C15}
\end{figure}

During the refitting of the ABOP, we made sure that the basic properties of iron in the bulk C15 phase given by DFT are well reproduced. Tab.~\ref{tab:C15bulk} shows the cohesive energies, lattice constants, and bulk moduli of the bcc ground state compared to bulk C15 iron. Additionally, we calculated the phonon dispersion of bulk C15 and compared with DFT results from Ref.~\cite{dezerald_stability_2014}, as shown in Fig.~\ref{fig:phonon_C15}. The fairly complex phonon dispersion with up to 18 branches is reproduced by the ABOP in overall good agreement with DFT, providing confidence that a realistic dynamical behaviour of large C15 clusters in $\alpha$-iron can be expected. In contrast, we found that bulk C15 in the EAM potentials is dynamically more unstable, and completely unstable in the AM04 potential (as evidenced by the appearance of acoustic phonon branches with imaginary frequencies).

\begin{figure}
 \centering
 \includegraphics[width=\linewidth]{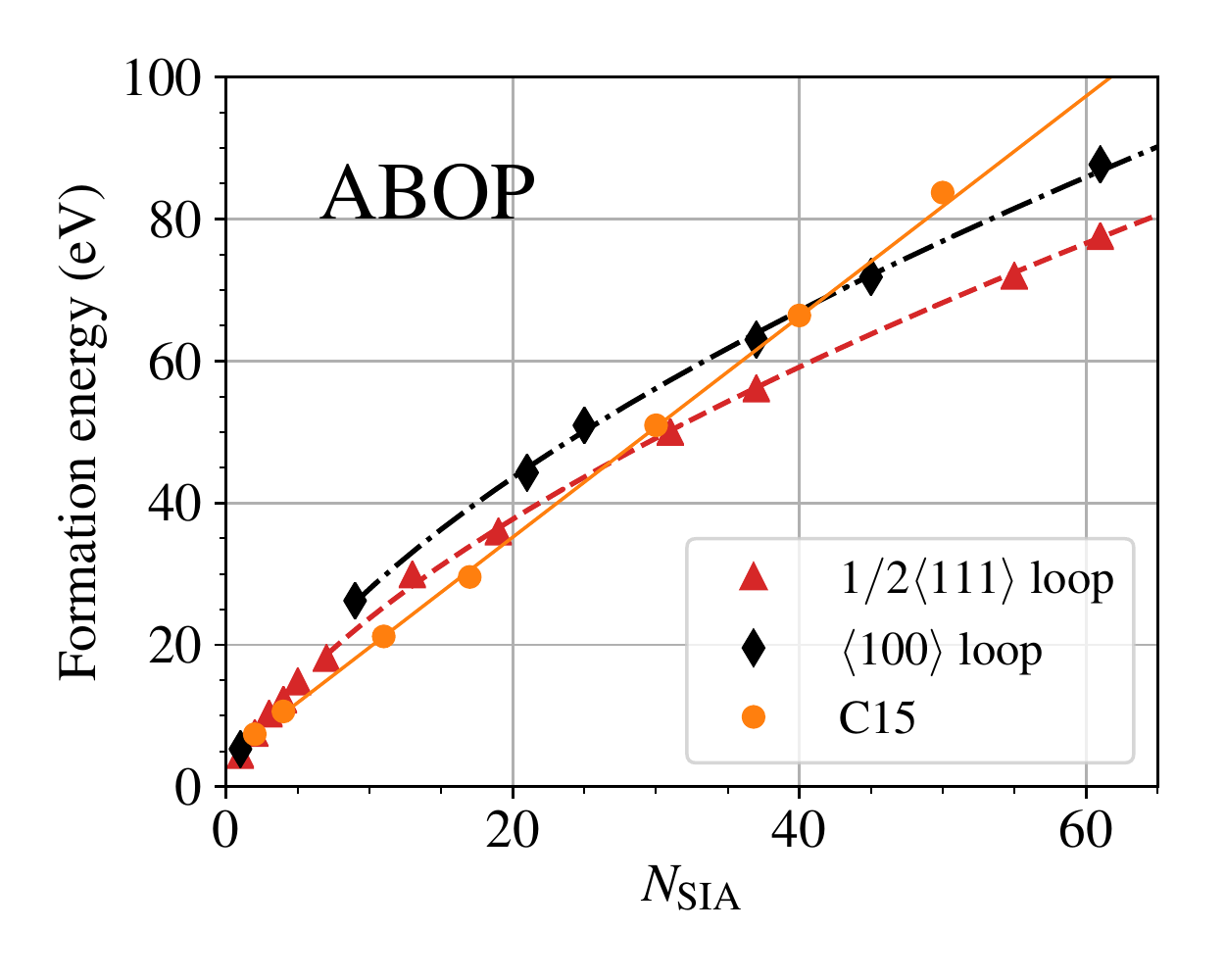}
 \caption{Formation energies of interstitial-type clusters in the ABOP.}
 \label{fig:Ef_C15loops}
\end{figure}

\begin{figure}
 \centering
 \includegraphics[width=\linewidth]{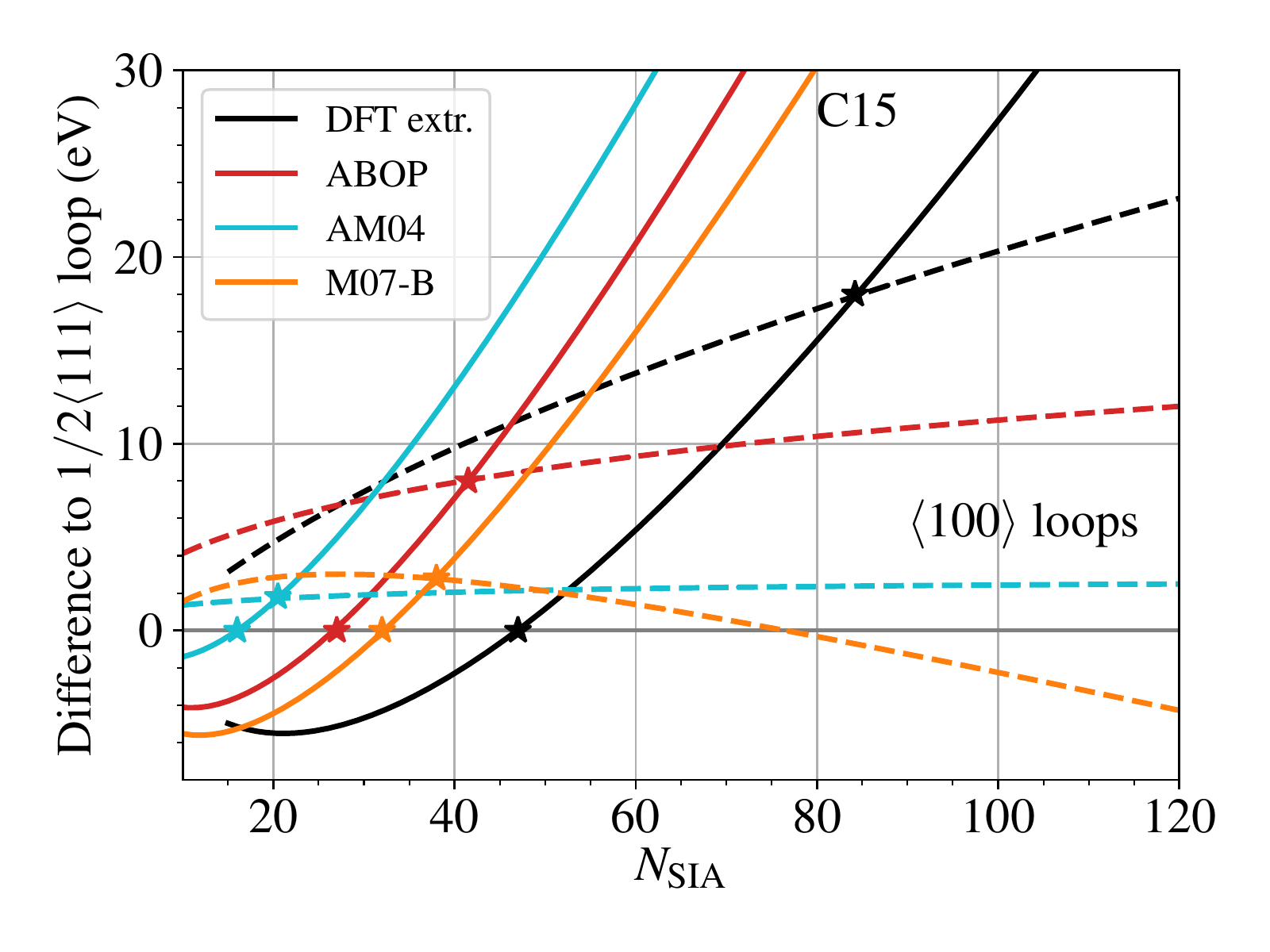}
 \caption{Difference in formation energy between C15 clusters (solid lines) and interstitial-type $\hkl<100>$ dislocation loops (dashed lines) compared to $1/2\hkl<111>$ loops (i.e. negative values means more stable than $1/2\hkl<111>$). Stars mark the stability cross-over points. The curves are plotted based on fits to the scaling laws of each cluster type. The DFT results are the scaling law fits from Ref.~\cite{alexander_ab_2016}.}
 \label{fig:111diff}
\end{figure}

For self-interstitial clusters in bcc iron, DFT predicts C15 clusters to energetically favoured over dislocation loops at small cluster sizes. At some critical size, $1/2\hkl<111>$ dislocation loops become lower in energy. At large cluster sizes, $1/2\hkl<111>$ are lowest in energy, followed by the $\hkl<100>$ loops and C15 clusters. This stability trend is qualitatively reproduced by the ABOP, as seen in Fig.~\ref{fig:Ef_C15loops} where formation energies of the three cluster types are plotted as functions of cluster size. However, like all EAM potentials, the crossovers in stability between the cluster types are underestimated. This is illustrated in Fig.~\ref{fig:111diff}, where the difference in formation energy between C15 and $1/2\hkl<111>$ loops, and between $\hkl<100>$ and $1/2\hkl<111>$ loops are plotted for the ABOP, the two EAM potentials, and the DFT-based scaling laws from Ref.~\cite{alexander_ab_2016}. The ABOP and EAM curves are plotted based on fits to the same scaling laws as the DFT data. As is clear from Fig.~\ref{fig:111diff}, the ABOP presents a critical improvement over the EAM potentials in terms of the relative stability of the two dislocation loops. $\hkl<100>$ loops are energetically very close to $1/2\hkl<111>$ loops in the EAM potentials, with the M07-B potential even predicting a crossover at around 80 SIAs, above which the $\hkl<100>$ loops are incorrectly lower in energy. Only the ABOP reproduces a clear difference in energy between the two dislocation loops. Nevertheless, it is worth noting that since theoretical work has shown that the increasingly anisotropic elasticity at higher temperatures eventually leads to $\hkl<100>$ loops being favoured~\cite{dudarev_effect_2008}, the EAM potentials can provide a qualitatively more realistic description at high temperatures, while the ABOP can be seen as more accurate at low temperatures.

\subsection{C15 cluster statistics in single cascades}
\label{sec:single_casc}

Having validated the revised ABOP for properties relevant for radiation damage, we use the potential to investigate the formation of C15 clusters in single collision cascades. In total, 150 cascades with the PKA energies  10, 20, and 50 keV were carried out (50 simulations per energy). Results from the EAM potentials were taken from Ref.~\cite{byggmastar_effects_2018} and further analysed. 10--18\% of the interstitial clusters were identified as C15-like clusters in the ABOP, 5--10\% in the M07-B potential, and 5--8\% in the AM04 potential, with no clear dependence on the PKA energy. Only clusters containing three or more interstitials were analysed (note however that the smallest C15 clusters contains only a net amount of two interstitials, but shows up as six interstitials and four vacancies after the Wigner-Seitz analysis, and were therefore also considered in the analysis). Previous simulations using the M07 potential reported that about 5\% of the clusters were C15-like~\cite{marinica_irradiation-induced_2012}, which is in line with our results with the M07-B potential (which only differ from the M07 potential at short interatomic distances~\cite{byggmastar_effects_2018}). Similar fractions can be expected at higher PKA energies, as cascade-splitting then becomes increasingly more likely and individual sub-cascades lead to similar cluster statistics.

\subsection{Growth and collapse of C15 clusters}
\label{sec:C15_stability}

\begin{table}
    \centering
    \begin{tabular}{llll}
        \toprule
        Potential & $1/2\hkl<111>$ & $\hkl<100>$ & mixed \\
        \midrule
        ABOP & 26/30 (87\%) & 1/30 (3\%) & 3/30 (10\%) \\
        M07-B & 9/20 (45\%) & 10/20 (50\%) & 1/20 (5\%) \\
        \bottomrule
    \end{tabular}
    \caption{Statistics of the resulting dislocation loop type after collapse of C15 clusters.}
    \label{tab:C15_collapse}
\end{table}

\begin{figure}
    \centering
    \includegraphics[width=\linewidth]{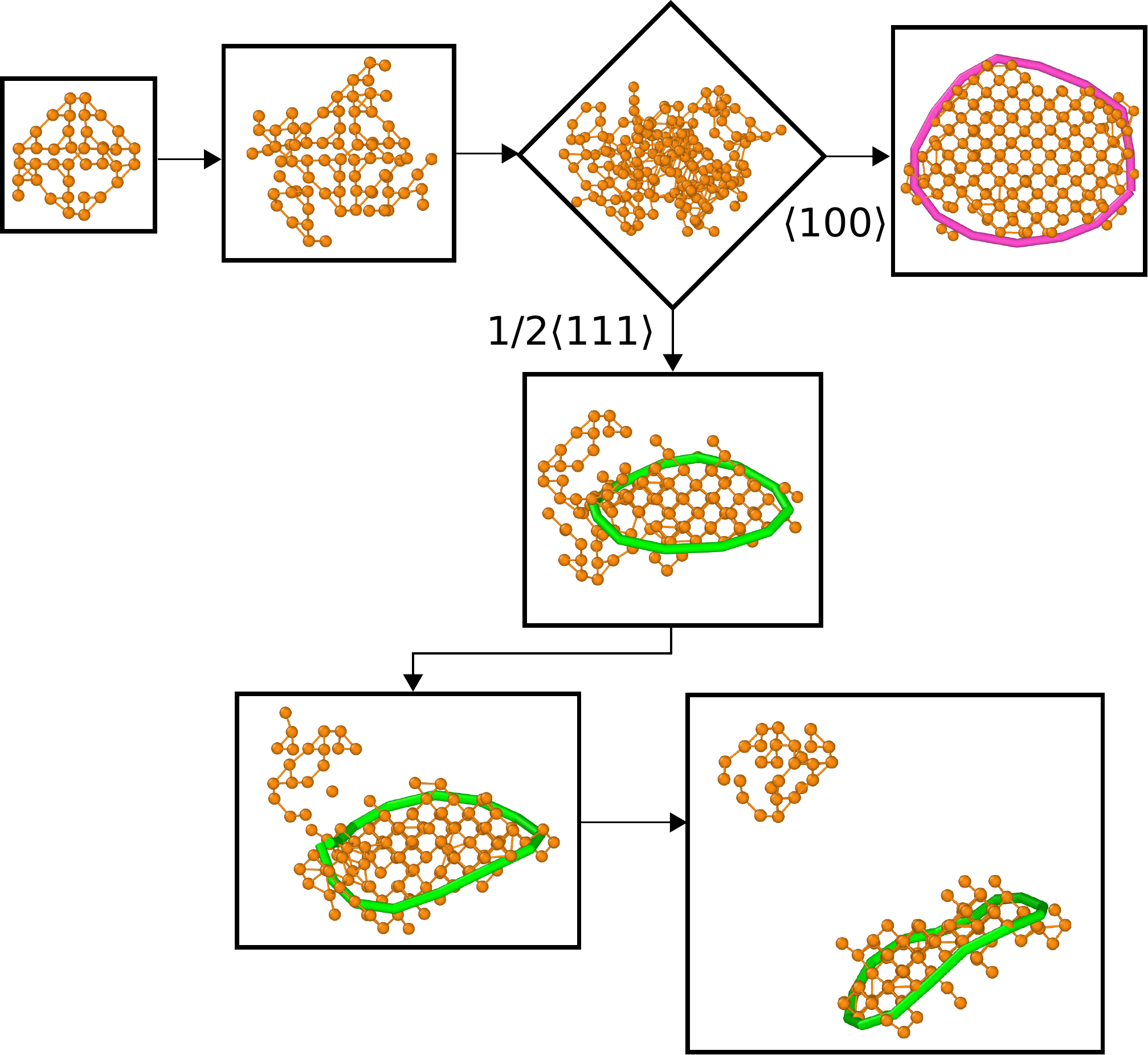}
    \caption{Schematic flowchart of the growth and collapse of a C15 cluster into a dislocation loop. Often when $1/2\hkl<111>$ loops are formed, a small C15 cluster is left behind as the loop is detached and migrates away.}
    \label{fig:C15collapse}
\end{figure}

Previous studies have suggested that once C15 clusters are formed, they can grow by absorbing migrating interstitials~\cite{marinica_irradiation-induced_2012,zhang_formation_2015,chartier_rearrangement_2019}. Zhang et al. used the M07 EAM potential to study the growth of C15 clusters by inserting single SIAs close to the cluster~\cite{zhang_formation_2015}. They observed that the growing C15 cluster eventually collapses into dislocation loops at sizes in the range 56--147 SIAs, with the most likely product being $\hkl<100>$ loops. Similarly, Chartier et al. recently observed, using the same interatomic potential, that the product of a collapsing C15 cluster depends on its size. A larger C15 cluster was seen to nucleate both \hkl<100> and $1/2\hkl<111>$ loops~\cite{chartier_rearrangement_2019}. As noted previously, however, the M07 potential incorrectly predicts the $\hkl<100>$ loops to be lower in energy at sizes above roughly 80 SIAs. Hence, it remains unclear whether collapse into $\hkl<100>$ loops is the most probable transformation path of large C15 clusters, or if these observations were due to an artefact of the used interatomic potential. We therefore carried out similar growth simulations using the ABOP, which gives the correct relative stabilities of the dislocation loops at all sizes.

In line with the results by Zhang et al.~\cite{zhang_formation_2015}, we found that the C15 clusters grow by capturing the nearby SIAs, and eventually become dynamically unstable and collapse into dislocation loops. Table.~\ref{tab:C15_collapse} summarises the statistics of the resulting dislocation loops. The vast majority of the collapsing C15 clusters transform into $1/2\hkl<111>$ loops in the ABOP. In contrast, the M07-B potential produces both loop types with roughly equal probability. The size at which the transformation into dislocation loops occurs, is highly stochastic. In the ABOP, in a few cases the C15 cluster collapsed into a loop already at 40--50 SIAs, while in some cases dislocation segments did not appear until cluster sizes above 120 SIAs. A similar size range was also observed for the M07-B potential, although the stochastic difference was smaller, with the transformation taking place at around 70--80 SIAs in the majority of the cases. In this size range, the \hkl<100> and $1/2\hkl<111>$ loops have almost identical formation energies in the M07-B potential (Fig.~\ref{fig:111diff}), which corresponds well to the observation that the probability of the C15 cluster transforming into either loop type is close to 50\%. In contrast, the ABOP favours $1/2\hkl<111>$ loops at all sizes, which leads to a very low probability of forming \hkl<100> loops from collapsing C15 clusters. Hence, the most likely outcome of collapsing C15 clusters directly correlates with the zero-kelvin formation energies of the two dislocation loops. This indicates that collapsing C15 clusters does not follow a transformation path that favours either loop type, and that instead it will in most cases simply restructure into the lowest-energy configuration. In a few cases, the transformation into dislocation loops did not complete, and the cluster remained as a dislocation segment terminating in a small stable C15 cluster even after growth to cluster sizes of 150 SIAs (these are labelled ''mixed'' in Tab.~\ref{tab:C15_collapse}). 

Fig.~\ref{fig:C15collapse} shows snapshots of the two possible transformation paths into dislocation loops. Interestingly, in the cases that resulted in $1/2\hkl<111>$ loops, the C15 cluster rarely transformed entirely into a loop. Instead, only part of the C15 cluster collapsed into a $1/2\hkl<111>$ loop, which is then eventually detached from the cluster while leaving a small C15 cluster behind as the loop rapidly migrates away. The remaining small C15 cluster could then continue growing, and eventually reach an unstable size again. In other words, collapse of growing C15 clusters does not necessarily mean that the C15 clusters will vanish. Instead, they can act as stationary nucleation sites and emitters of $1/2\hkl<111>$ loops.

In the ABOP, only one case out of 30 resulted in a $\hkl<100>$ loop. Our results therefore confirm that C15 clusters can collapse into $\hkl<100>$ loops, but that it is not very likely at low temperatures. However, as mentioned previously, it is known from both theoretical work~\cite{dudarev_effect_2008} and experimental observations~\cite{yao_temperature_2010} that $\hkl<100>$ loops become more stable as the temperature increases. It is, therefore, reasonable to assume that the frequent formation of $\hkl<100>$ loops from unstable C15 clusters in the M07-B potential is more representative of high-temperature ($>600$ K~\cite{dudarev_effect_2008}) irradiation (regardless of the temperature used in the simulations), while the results from the ABOP can be assumed to be more valid for lower temperatures.

The low probability of collapse into \hkl<100> loops predicted by the ABOP can be justified based on recent Object kinetic Monte Carlo (OKMC) simulations compared with experimental results. Balbuena et al. studied the population of \hkl<100> loops in a thin foil using OKMC, with the assumption that all \hkl<100> loops formed from formation and collapse of C15 clusters~\cite{balbuena_insights_2019}. They assumed that a certain fraction of small clusters transform into C15, and a fraction of these subsequently collapsed into \hkl<100> loops. Good agreement with experiments was obtained when assuming that the total probability for this \hkl<100> nucleation sequence is 0.1\%. The corresponding \hkl<100> nucleation probabilities can be estimated based on our MD simulations. The ABOP predicts 10--18\% of clusters formed in cascades to be C15 clusters, and that upon growth 3.33\% (1/30) of them collapse into \hkl<100> loops, yielding a nucleation probability in the range 0.33--0.6\%. Considering the poor statistics of the growth simulations (1/30), the uncertainty of this probability is large, and is therefore roughly consistent with the 0.1\% empirical estimate by Balbuena et al.~\cite{balbuena_insights_2019}. In contrast, the corresponding nucleation probability in the M07-B potential is 2.5--5\%, indicating an overestimation of \hkl<100> loops compared to experiments.

\subsection{Accumulation of C15 clusters under continuous irradiation}
\label{sec:C15_stats}

\begin{figure}
 \centering
 \includegraphics[width=\linewidth]{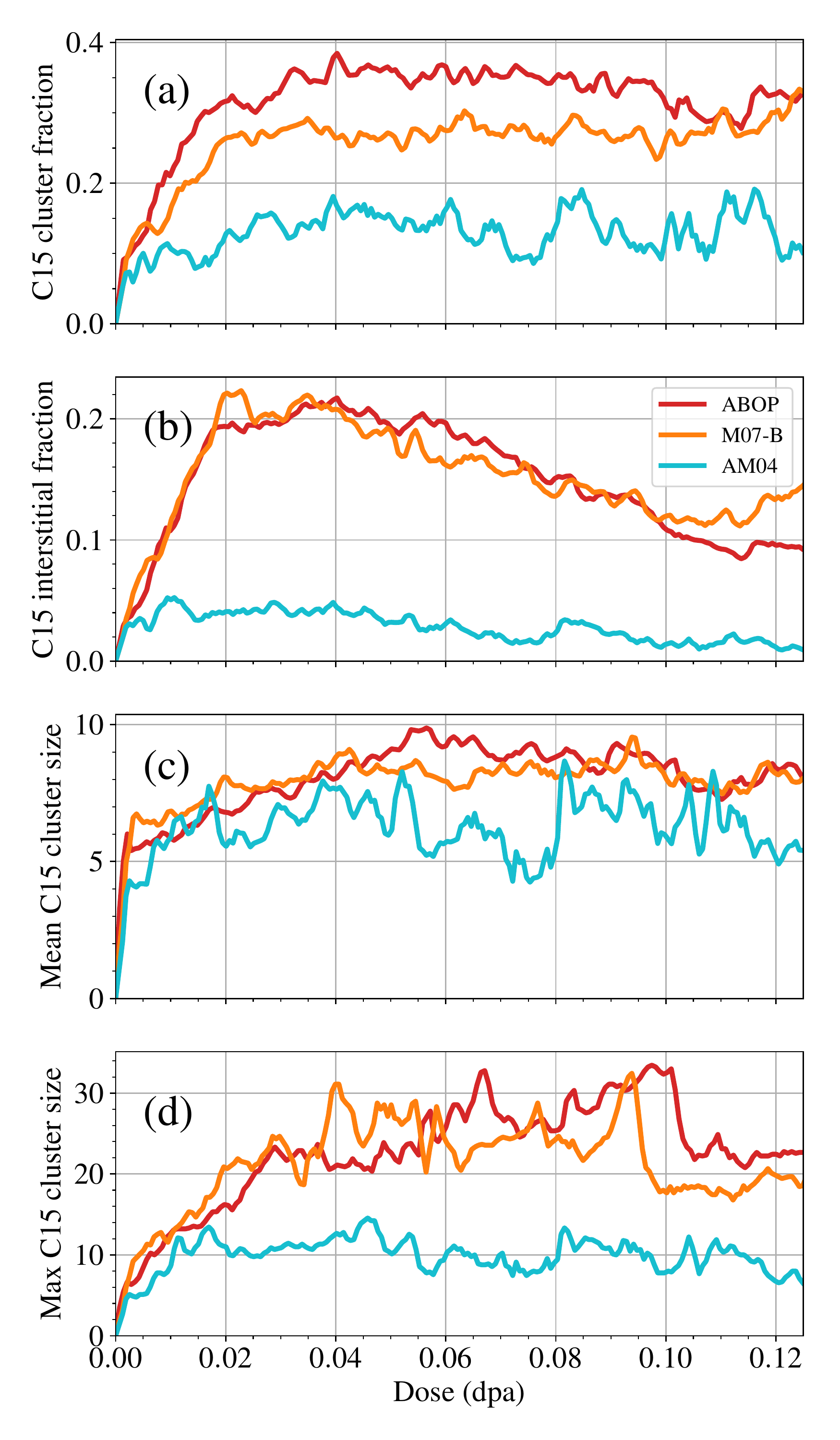}
 \caption{(a): Fractions of interstitial clusters identified as C15-like clusters as a function of dose (note that these include C15 clusters attached to dislocation loops). (b): Fractions of interstitials contained in C15-like clusters. (c): Average size of C15 clusters. (d): Maximum size of C15 clusters.}
 \label{fig:C15-dose}
\end{figure}

\begin{figure}
    \centering
    \includegraphics[width=0.48\linewidth]{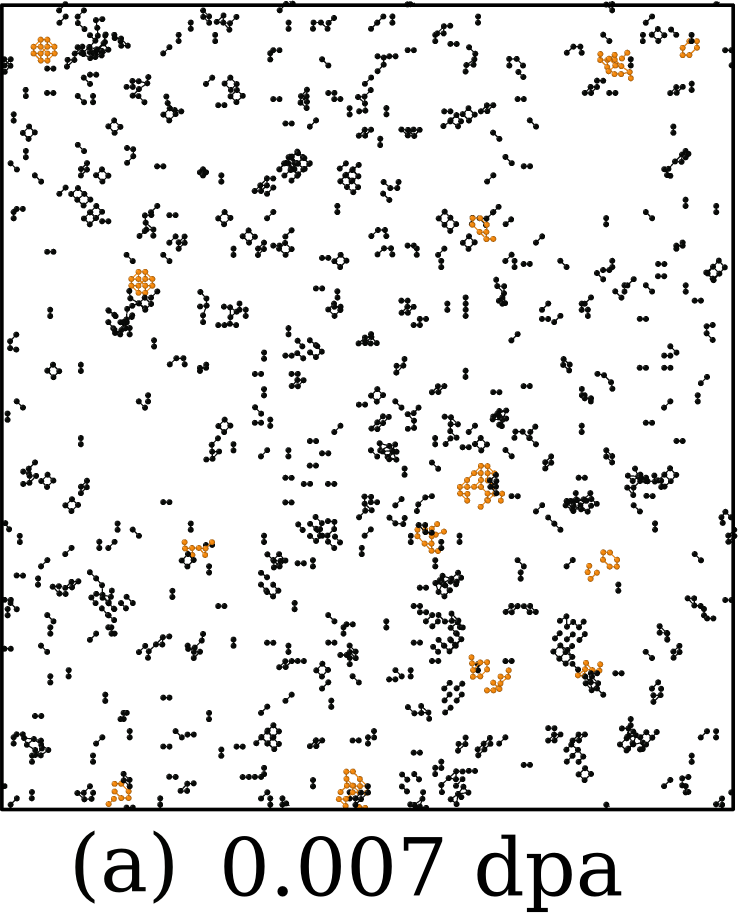}
    \includegraphics[width=0.48\linewidth]{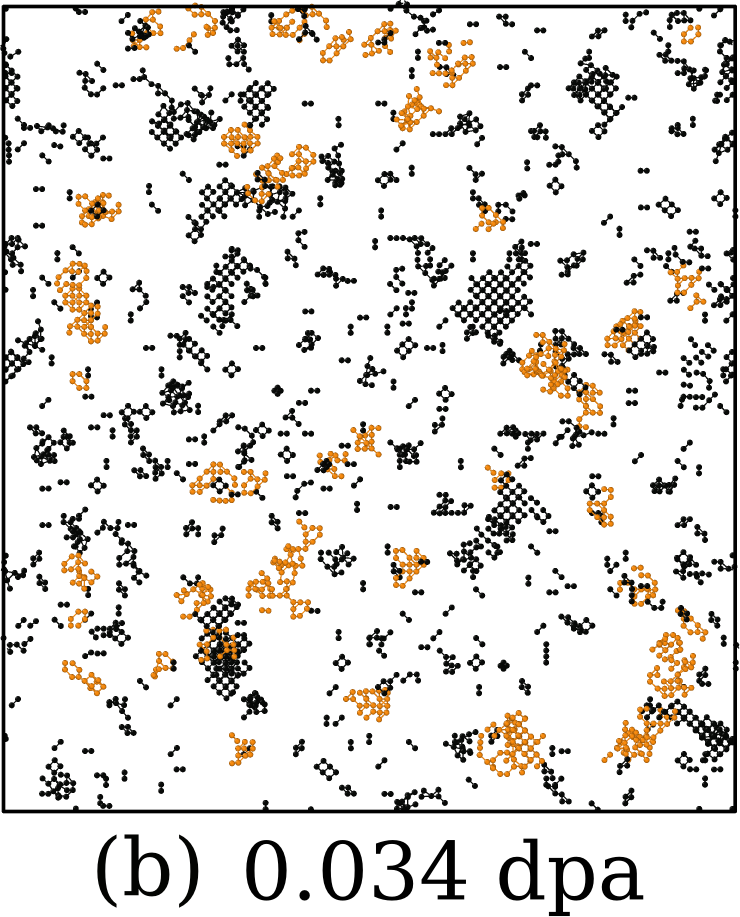}
    \includegraphics[width=0.48\linewidth]{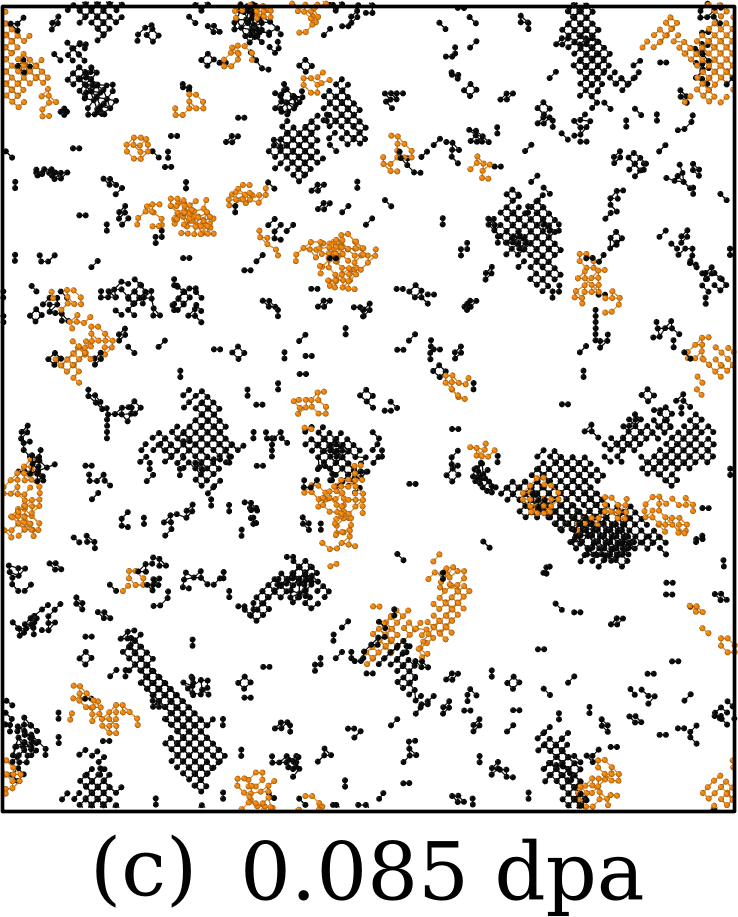}
    \includegraphics[width=0.48\linewidth]{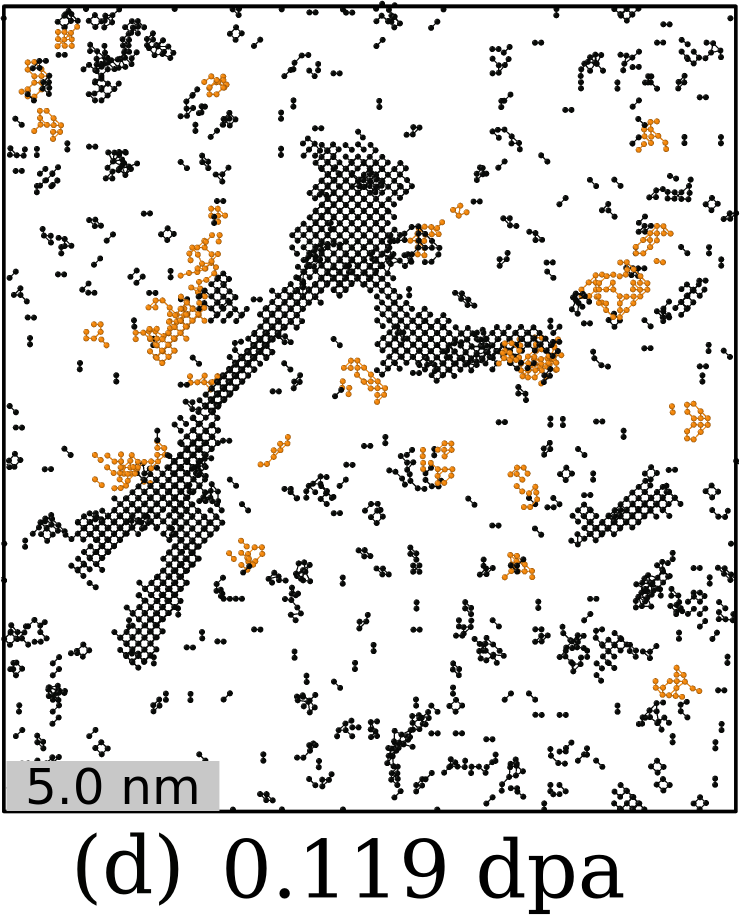}
    \caption{Interstitial atoms accumulated at different doses in cascade simulations using the ABOP. C15-like clusters are coloured orange and all other interstitials are black. Dislocation loops are visible as large black interstitial clusters.}
    \label{fig:C15clusters}
\end{figure}

In the previous sections we have confirmed that C15 clusters can form directly as a result of a collision cascade, and that they are energetically and dynamically stable up to a critical size, after which they collapse into dislocation loops. In this section, we observe the entire life cycle of C15 clusters in continuous irradiation, from formation to collapse or absorption by a larger cluster. This is similar to the recent work by Chartier et al., where electron irradiation was mimicked by inserting Frenkel pairs at regular intervals~\cite{chartier_rearrangement_2019}. Our work deals with overlapping cascades induced by recoils from neutron or ion irradiation at low temperatures where thermal migration is negligible (due to the time scale limitation of MD). We analyse the population of C15 clusters in simulations of 2000 cumulative 5 keV collision cascades. New simulations are carried out with the ABOP, and the results are compared with further analysis of our previous simulation data from Ref.~\cite{byggmastar_effects_2018} using the EAM potentials.

Fig.~\ref{fig:C15-dose}a shows the fraction of clusters that are identified as C15-like as a function of dose in the three different interatomic potentials. The data are averaged over three separate simulation runs. The AM04 potential shows by far the lowest fractions of C15 clusters, as is expected since the stability of C15 clusters compared to dislocation loops in this potential is strongly underestimated (Fig.~\ref{fig:111diff}). Both the ABOP and the M07-B potential show a saturation of around 30\% C15 clusters out of all interstitial clusters containing more than two interstitial atoms. The identified cluster fractions are, however, slightly misleading for two reasons. First, dislocation loops are frequently attached to C15 clusters, in which case the entire cluster is here counted as one C15-like cluster. Second, as the dose increases, the mobile $1/2\hkl<111>$ loops are rapidly growing to large sizes as they absorb small clusters, resulting in a rapid decrease in the total number of clusters. Therefore, in Fig.~\ref{fig:C15-dose}b we show the fraction of interstitial atoms identified as belonging to a C15-like geometry. These fractions show that after reaching a maximum C15 population at around 0.02--0.05 dpa, the fraction of C15-like interstitials starts decreasing until they start saturating at around 10\% in the ABOP and M07-B potentials. Both the ABOP and the M07-B potentials show the same trend.

Figs.~\ref{fig:C15-dose}c-d show the mean and maximum sizes of the C15 clusters as a function of dose. The majority of C15 clusters are small at all doses, with the average size between 5 and 10 interstitials in all potentials. The maximum size of any C15 cluster reaches about 30 interstitials in the ABOP and M07-B potentials, and around 10 interstitials in the AM04 potential. This corresponds to diameters in the 1--2 nm range. The maximum sizes are very close to the cross-overs in energy between C15 and $1/2 \hkl<111>$ loops (Fig.~\ref{fig:111diff}). The maximum sizes are much smaller than observed in the controlled growth simulations, which is due to the continuous irradiation and presence of larger mobile dislocation loops resulting in frequent cluster transformations due to cascade overlap, and absorption of smaller clusters. These effects are, however, severely overestimated due to the rapid dose rate required in MD simulations, and in experimental conditions the C15 clusters would likely grow more undisturbed towards larger sizes. Additionally, since all potentials underestimate the cross-over in energy between C15 clusters and loops, the size range of C15 clusters observed here are likely underestimated accordingly. We also note that overall the ABOP and M07-B potentials show remarkably similar results in Fig.~\ref{fig:C15-dose}, owing to the similar predicted stability of C15 clusters.

The evolution of the C15 cluster statistics shown in Fig.~\ref{fig:C15-dose} can be understood by visually analysing the defect structure at different doses. Fig.~\ref{fig:C15clusters} shows snapshots of the interstitial atoms at different doses in the ABOP, with C15-like clusters coloured orange and other interstitials black. C15 clusters are frequently forming already at very low doses (Fig.~\ref{fig:C15clusters}a). The C15 population is largely controlled by the presence and growth of mobile $1/2\hkl<111>$ loops. At doses below 0.05 dpa (Fig.~\ref{fig:C15clusters}b), the clusters are still relatively small, and in the range where C15 clusters are energetically favoured. As the clusters absorb more interstitials, $1/2\hkl<111>$ loops eventually start dominating and absorbing other small clusters, including C15 clusters, as they migrate rapidly through the simulation box. This is visible in Fig.~\ref{fig:C15-dose}b as a stable decrease in interstitial fraction and is illustrated in Fig.~\ref{fig:C15clusters}c. At higher doses, the C15 population has reached an equilibrium where new clusters are formed roughly at the same rate as they are absorbed by the large $1/2\hkl<111>$ loops, corresponding to Fig.~\ref{fig:C15clusters}d.

\section{Summary and conclusions}
\label{sec:concl}

Using molecular dynamics simulations and a new interatomic bond-order potential that reproduces the correct relative stability of C15 clusters, \hkl<100> loops and $1/2\hkl<111>$ loops, we investigated the formation, growth, collapse, and population of self-interstitial C15 Laves phase clusters in $\alpha$-iron. We found that about 5--20\% of the interstitial clusters formed in collision cascades are C15-like clusters. Furthermore, we observed that when growing C15 clusters eventually collapse, $1/2\hkl<111>$ loops is the predominant product at least at low temperatures. During collapse of C15 into $1/2\hkl<111>$ loops, the loop is often detached and escapes before a full transformation is complete, leaving behind a small stable C15 cluster. Hence, C15 clusters can act as long-lived and stationary emitters of $1/2\hkl<111>$ loops. Our results also show that C15 clusters can in rare occasions collapse into $\hkl<100>$ loops, in line with earlier studies, and that this is likely more common at higher temperatures. Under rapid continuous irradiation, up to 30\% of the self-interstitial clusters or 10--20\% of the total number of self-interstitial atoms were identified as C15-like. The majority of C15 clusters remain small in size, in the range of 1--2 nm. We conclude that the frequent formation and large populations of the small and immobile C15 clusters can have a significant impact on the microstructural evolution of iron-based materials.

\section*{Acknowledgements}

This work has been carried out within the framework of the EUROfusion Consortium and has received funding from the Euratom research and training programme 2014-2018 and 2019-2020 under grant agreement No 633053. The views and opinions expressed herein do not necessarily reflect those of the European Commission. Grants of computer capacity from CSC - IT Center for Science, Finland, as well as from the Finnish Grid and Cloud Infrastructure (persistent identifier urn:nbn:fi:research-infras-2016072533) are gratefully acknowledged.

\section*{Data Availability}

The raw data required to reproduce these findings are available upon request from the authors. The processed data required to reproduce these findings are available upon request from the authors.

\section*{References}
\bibliography{mybib}

\begin{thebibliography}{22}
\expandafter\ifx\csname natexlab\endcsname\relax\def\natexlab#1{#1}\fi
\providecommand{\url}[1]{\texttt{#1}}
\providecommand{\href}[2]{#2}
\providecommand{\path}[1]{#1}
\providecommand{\DOIprefix}{doi:}
\providecommand{\ArXivprefix}{arXiv:}
\providecommand{\URLprefix}{URL: }
\providecommand{\Pubmedprefix}{pmid:}
\providecommand{\doi}[1]{\href{http://dx.doi.org/#1}{\path{#1}}}
\providecommand{\Pubmed}[1]{\href{pmid:#1}{\path{#1}}}
\providecommand{\bibinfo}[2]{#2}
\ifx\xfnm\relax \def\xfnm[#1]{\unskip,\space#1}\fi
%Type = Article
\bibitem[{Marinica et~al.(2012)Marinica, Willaime, and
  Crocombette}]{marinica_irradiation-induced_2012}
\bibinfo{author}{M.-C. Marinica}, \bibinfo{author}{F.~Willaime},
  \bibinfo{author}{J.-P. Crocombette}, \bibinfo{journal}{Phys. Rev. Lett.}
  \bibinfo{volume}{108} (\bibinfo{year}{2012}) \bibinfo{pages}{025501}.
  \DOIprefix\doi{10.1103/PhysRevLett.108.025501}.
%Type = Article
\bibitem[{Byggm\"astar et~al.(2018)Byggm\"astar, Granberg, and
  Nordlund}]{byggmastar_effects_2018}
\bibinfo{author}{J.~Byggm\"astar}, \bibinfo{author}{F.~Granberg},
  \bibinfo{author}{K.~Nordlund}, \bibinfo{journal}{Journal of Nuclear
  Materials} \bibinfo{volume}{508} (\bibinfo{year}{2018})
  \bibinfo{pages}{530--539}. \DOIprefix\doi{10.1016/j.jnucmat.2018.06.005}.
%Type = Article
\bibitem[{Chartier and Marinica(2019)}]{chartier_rearrangement_2019}
\bibinfo{author}{A.~Chartier}, \bibinfo{author}{M.~C. Marinica},
  \bibinfo{journal}{Acta Materialia} \bibinfo{volume}{180}
  (\bibinfo{year}{2019}) \bibinfo{pages}{141--148}.
  \DOIprefix\doi{10.1016/j.actamat.2019.09.007}.
%Type = Article
\bibitem[{Yi et~al.(2013)Yi, Jenkins, Briceno, Roberts, Zhou, and
  Kirk}]{yi_situ_2013}
\bibinfo{author}{X.~Yi}, \bibinfo{author}{M.~L. Jenkins},
  \bibinfo{author}{M.~Briceno}, \bibinfo{author}{S.~G. Roberts},
  \bibinfo{author}{Z.~Zhou}, \bibinfo{author}{M.~A. Kirk},
  \bibinfo{journal}{Philos. Mag.} \bibinfo{volume}{93} (\bibinfo{year}{2013})
  \bibinfo{pages}{1715--1738}. \DOIprefix\doi{10.1080/14786435.2012.754110}.
%Type = Article
\bibitem[{Yi et~al.(2015)Yi, Jenkins, Hattar, Edmondson, and
  Roberts}]{yi_characterisation_2015}
\bibinfo{author}{X.~Yi}, \bibinfo{author}{M.~L. Jenkins},
  \bibinfo{author}{K.~Hattar}, \bibinfo{author}{P.~D. Edmondson},
  \bibinfo{author}{S.~G. Roberts}, \bibinfo{journal}{Acta Materialia}
  \bibinfo{volume}{92} (\bibinfo{year}{2015}) \bibinfo{pages}{163--177}.
  \DOIprefix\doi{10.1016/j.actamat.2015.04.015}.
%Type = Article
\bibitem[{Sand et~al.(2013)Sand, Dudarev, and Nordlund}]{sand_high-energy_2013}
\bibinfo{author}{A.~E. Sand}, \bibinfo{author}{S.~L. Dudarev},
  \bibinfo{author}{K.~Nordlund}, \bibinfo{journal}{EPL} \bibinfo{volume}{103}
  (\bibinfo{year}{2013}) \bibinfo{pages}{46003}.
  \DOIprefix\doi{10.1209/0295-5075/103/46003}.
%Type = Article
\bibitem[{Byggm\"astar et~al.(2019)Byggm\"astar, Granberg, Sand, Pirttikoski,
  Alexander, Marinica, and Nordlund}]{byggmastar_collision_2019}
\bibinfo{author}{J.~Byggm\"astar}, \bibinfo{author}{F.~Granberg},
  \bibinfo{author}{A.~E. Sand}, \bibinfo{author}{A.~Pirttikoski},
  \bibinfo{author}{R.~Alexander}, \bibinfo{author}{M.-C. Marinica},
  \bibinfo{author}{K.~Nordlund}, \bibinfo{journal}{J. Phys.: Condens. Matter}
  \bibinfo{volume}{31} (\bibinfo{year}{2019}) \bibinfo{pages}{245402}.
  \DOIprefix\doi{10.1088/1361-648X/ab0682}.
%Type = Article
\bibitem[{Alexander et~al.(2016)Alexander, Marinica, Proville, Willaime,
  Arakawa, Gilbert, and Dudarev}]{alexander_ab_2016}
\bibinfo{author}{R.~Alexander}, \bibinfo{author}{M.-C. Marinica},
  \bibinfo{author}{L.~Proville}, \bibinfo{author}{F.~Willaime},
  \bibinfo{author}{K.~Arakawa}, \bibinfo{author}{M.~R. Gilbert},
  \bibinfo{author}{S.~L. Dudarev}, \bibinfo{journal}{Phys. Rev. B}
  \bibinfo{volume}{94} (\bibinfo{year}{2016}) \bibinfo{pages}{024103}.
  \DOIprefix\doi{10.1103/PhysRevB.94.024103}.
%Type = Article
\bibitem[{Malerba et~al.(2010)Malerba, Marinica, Anento, Bj\"orkas, Nguyen,
  Domain, Djurabekova, Olsson, Nordlund, Serra, Terentyev, Willaime, and
  Becquart}]{malerba_comparison_2010}
\bibinfo{author}{L.~Malerba}, \bibinfo{author}{M.~C. Marinica},
  \bibinfo{author}{N.~Anento}, \bibinfo{author}{C.~Bj\"orkas},
  \bibinfo{author}{H.~Nguyen}, \bibinfo{author}{C.~Domain},
  \bibinfo{author}{F.~Djurabekova}, \bibinfo{author}{P.~Olsson},
  \bibinfo{author}{K.~Nordlund}, \bibinfo{author}{A.~Serra},
  \bibinfo{author}{D.~Terentyev}, \bibinfo{author}{F.~Willaime},
  \bibinfo{author}{C.~S. Becquart}, \bibinfo{journal}{Journal of Nuclear
  Materials} \bibinfo{volume}{406} (\bibinfo{year}{2010})
  \bibinfo{pages}{19--38}. \DOIprefix\doi{10.1016/j.jnucmat.2010.05.017}.
%Type = Article
\bibitem[{Zhang et~al.(2015)Zhang, Bai, Tonks, and
  Biner}]{zhang_formation_2015}
\bibinfo{author}{Y.~Zhang}, \bibinfo{author}{X.-M. Bai}, \bibinfo{author}{M.~R.
  Tonks}, \bibinfo{author}{S.~B. Biner}, \bibinfo{journal}{Scripta Materialia}
  \bibinfo{volume}{98} (\bibinfo{year}{2015}) \bibinfo{pages}{5--8}.
  \DOIprefix\doi{10.1016/j.scriptamat.2014.10.033}.
%Type = Article
\bibitem[{M\"uller et~al.(2007)M\"uller, Erhart, and
  Albe}]{muller_analytic_2007}
\bibinfo{author}{M.~M\"uller}, \bibinfo{author}{P.~Erhart},
  \bibinfo{author}{K.~Albe}, \bibinfo{journal}{J. Phys.: Condens. Matter}
  \bibinfo{volume}{19} (\bibinfo{year}{2007}) \bibinfo{pages}{326220}.
  \DOIprefix\doi{10.1088/0953-8984/19/32/326220}.
%Type = Article
\bibitem[{Bj\"orkas and Nordlund(2007)}]{bjorkas_comparative_2007}
\bibinfo{author}{C.~Bj\"orkas}, \bibinfo{author}{K.~Nordlund},
  \bibinfo{journal}{Nuclear Instruments and Methods in Physics Research Section
  B: Beam Interactions with Materials and Atoms} \bibinfo{volume}{259}
  (\bibinfo{year}{2007}) \bibinfo{pages}{853--860}.
  \DOIprefix\doi{10.1016/j.nimb.2007.03.076}.
%Type = Article
\bibitem[{D\'ezerald et~al.(2014)D\'ezerald, Marinica, Ventelon, Rodney, and
  Willaime}]{dezerald_stability_2014}
\bibinfo{author}{L.~D\'ezerald}, \bibinfo{author}{M.~C. Marinica},
  \bibinfo{author}{L.~Ventelon}, \bibinfo{author}{D.~Rodney},
  \bibinfo{author}{F.~Willaime}, \bibinfo{journal}{Journal of Nuclear
  Materials} \bibinfo{volume}{449} (\bibinfo{year}{2014})
  \bibinfo{pages}{219--224}. \DOIprefix\doi{10.1016/j.jnucmat.2014.02.012}.
%Type = Article
\bibitem[{Plimpton(1995)}]{plimpton_fast_1995}
\bibinfo{author}{S.~Plimpton}, \bibinfo{journal}{Journal of Computational
  Physics} \bibinfo{volume}{117} (\bibinfo{year}{1995}) \bibinfo{pages}{1--19}.
  \DOIprefix\doi{10.1006/jcph.1995.1039},
  \bibinfo{note}{http://lammps.sandia.gov}.
%Type = Article
\bibitem[{Larsen et~al.(2017)Larsen, Mortensen, Blomqvist, Castelli,
  Christensen, {Marcin Du\l{}ak}, Friis, Groves, Hammer, Hargus, Hermes,
  Jennings, Jensen, Kermode, Kitchin, Kolsbjerg, Kubal, {Kristen Kaasbjerg},
  Lysgaard, Maronsson, Maxson, Olsen, Pastewka, {Andrew Peterson}, Rostgaard,
  Schi\o{}tz, Sch\"utt, Strange, Thygesen, {Tejs Vegge}, Vilhelmsen, Walter,
  Zeng, and Jacobsen}]{larsen_atomic_2017}
\bibinfo{author}{A.~H. Larsen}, \bibinfo{author}{J.~J. Mortensen},
  \bibinfo{author}{J.~Blomqvist}, \bibinfo{author}{I.~E. Castelli},
  \bibinfo{author}{R.~Christensen}, \bibinfo{author}{{Marcin Du\l{}ak}},
  \bibinfo{author}{J.~Friis}, \bibinfo{author}{M.~N. Groves},
  \bibinfo{author}{B.~Hammer}, \bibinfo{author}{C.~Hargus},
  \bibinfo{author}{E.~D. Hermes}, \bibinfo{author}{P.~C. Jennings},
  \bibinfo{author}{P.~B. Jensen}, \bibinfo{author}{J.~Kermode},
  \bibinfo{author}{J.~R. Kitchin}, \bibinfo{author}{E.~L. Kolsbjerg},
  \bibinfo{author}{J.~Kubal}, \bibinfo{author}{{Kristen Kaasbjerg}},
  \bibinfo{author}{S.~Lysgaard}, \bibinfo{author}{J.~B. Maronsson},
  \bibinfo{author}{T.~Maxson}, \bibinfo{author}{T.~Olsen},
  \bibinfo{author}{L.~Pastewka}, \bibinfo{author}{{Andrew Peterson}},
  \bibinfo{author}{C.~Rostgaard}, \bibinfo{author}{J.~Schi\o{}tz},
  \bibinfo{author}{O.~Sch\"utt}, \bibinfo{author}{M.~Strange},
  \bibinfo{author}{K.~S. Thygesen}, \bibinfo{author}{{Tejs Vegge}},
  \bibinfo{author}{L.~Vilhelmsen}, \bibinfo{author}{M.~Walter},
  \bibinfo{author}{Z.~Zeng}, \bibinfo{author}{K.~W. Jacobsen},
  \bibinfo{journal}{J. Phys.: Condens. Matter} \bibinfo{volume}{29}
  (\bibinfo{year}{2017}) \bibinfo{pages}{273002}.
  \DOIprefix\doi{10.1088/1361-648X/aa680e}.
%Type = Article
\bibitem[{Nordlund(1995)}]{nordlund_molecular_1995}
\bibinfo{author}{K.~Nordlund}, \bibinfo{journal}{Computational Materials
  Science} \bibinfo{volume}{3} (\bibinfo{year}{1995})
  \bibinfo{pages}{448--456}. \DOIprefix\doi{10.1016/0927-0256(94)00085-Q}.
%Type = Article
\bibitem[{Nordlund et~al.(1998)Nordlund, Ghaly, Averback, Caturla, {Diaz de la
  Rubia}, and Tarus}]{nordlund_defect_1998}
\bibinfo{author}{K.~Nordlund}, \bibinfo{author}{M.~Ghaly},
  \bibinfo{author}{R.~S. Averback}, \bibinfo{author}{M.~Caturla},
  \bibinfo{author}{T.~{Diaz de la Rubia}}, \bibinfo{author}{J.~Tarus},
  \bibinfo{journal}{Phys. Rev. B} \bibinfo{volume}{57} (\bibinfo{year}{1998})
  \bibinfo{pages}{7556--7570}. \DOIprefix\doi{10.1103/PhysRevB.57.7556}.
%Type = Article
\bibitem[{Ackland et~al.(2004)Ackland, Mendelev, Srolovitz, Han, and
  Barashev}]{ackland_development_2004}
\bibinfo{author}{G.~J. Ackland}, \bibinfo{author}{M.~I. Mendelev},
  \bibinfo{author}{D.~J. Srolovitz}, \bibinfo{author}{S.~Han},
  \bibinfo{author}{A.~V. Barashev}, \bibinfo{journal}{J. Phys.: Condens.
  Matter} \bibinfo{volume}{16} (\bibinfo{year}{2004}) \bibinfo{pages}{S2629}.
  \DOIprefix\doi{10.1088/0953-8984/16/27/003}.
%Type = Book
\bibitem[{Haynes(2015)}]{haynes_crc_2015}
\bibinfo{author}{W.~M. Haynes}, \bibinfo{title}{{{CRC Handbook}} of
  {{Chemistry}} and {{Physics}}}, \bibinfo{edition}{96th} ed.,
  \bibinfo{publisher}{{CRC Press}}, \bibinfo{address}{Boca Raton, FL},
  \bibinfo{year}{2015}.
%Type = Article
\bibitem[{Dudarev et~al.(2008)Dudarev, Bullough, and
  Derlet}]{dudarev_effect_2008}
\bibinfo{author}{S.~L. Dudarev}, \bibinfo{author}{R.~Bullough},
  \bibinfo{author}{P.~M. Derlet}, \bibinfo{journal}{Phys. Rev. Lett.}
  \bibinfo{volume}{100} (\bibinfo{year}{2008}) \bibinfo{pages}{135503}.
  \DOIprefix\doi{10.1103/PhysRevLett.100.135503}.
%Type = Article
\bibitem[{Yao et~al.(2010)Yao, Jenkins, {Hern\'andez-Mayoral}, and
  Kirk}]{yao_temperature_2010}
\bibinfo{author}{Z.~Yao}, \bibinfo{author}{M.~L. Jenkins},
  \bibinfo{author}{M.~{Hern\'andez-Mayoral}}, \bibinfo{author}{M.~A. Kirk},
  \bibinfo{journal}{Philos. Mag.} \bibinfo{volume}{90} (\bibinfo{year}{2010})
  \bibinfo{pages}{4623--4634}. \DOIprefix\doi{10.1080/14786430903430981}.
%Type = Article
\bibitem[{Balbuena et~al.(2019)Balbuena, Aliaga, Dopico,
  {Hern{\'a}ndez-Mayoral}, Malerba, {Martin-Bragado}, and
  Caturla}]{balbuena_insights_2019}
\bibinfo{author}{J.~P. Balbuena}, \bibinfo{author}{M.~J. Aliaga},
  \bibinfo{author}{I.~Dopico}, \bibinfo{author}{M.~{Hern{\'a}ndez-Mayoral}},
  \bibinfo{author}{L.~Malerba}, \bibinfo{author}{I.~{Martin-Bragado}},
  \bibinfo{author}{M.~J. Caturla}, \bibinfo{journal}{Journal of Nuclear
  Materials} \bibinfo{volume}{521} (\bibinfo{year}{2019})
  \bibinfo{pages}{71--80}. \DOIprefix\doi{10.1016/j.jnucmat.2019.04.030}.

\end{thebibliography}


\providecommand{\newblock}{}
\begin{thebibliography}{10}
\expandafter\ifx\csname url\endcsname\relax
  \def\url#1{{\tt #1}}\fi
\expandafter\ifx\csname urlprefix\endcsname\relax\def\urlprefix{URL }\fi
\providecommand{\eprint}[2][]{\url{#2}}

\bibitem{albe_modeling_2002-1}
Albe K, Nordlund K and Averback R~S 2002 {\em Phys. Rev. B\/} {\bf 65} 195124

\bibitem{byggmastar_analytical_2019}
Byggm\"astar J, Nagel M, Albe K, Henriksson K~O~E and Nordlund K 2019 {\em J.
  Phys.: Condens. Matter\/} {\bf 31} 215401 ISSN 0953-8984

\bibitem{ziegler_stopping_1985}
Ziegler J~F, Biersack J~P and Littmarck U 1985 The {{Stopping}} and {{Range}}
  of {{Ions}} in {{Matter}} {\em Treatise on {{Heavy}}-{{Ion Science}}\/} (New
  York: {Pergamon}) pp 93--129 ISBN 978-1-4615-8105-5 978-1-4615-8103-1

\bibitem{muller_analytic_2007}
M\"uller M, Erhart P and Albe K 2007 {\em J. Phys.: Condens. Matter\/} {\bf 19}
  326220 ISSN 0953-8984

\bibitem{bjorkas_comparative_2007}
Bj\"orkas C and Nordlund K 2007 {\em Nuclear Instruments and Methods in Physics
  Research Section B: Beam Interactions with Materials and Atoms\/} {\bf 259}
  853--860 ISSN 0168-583X

\bibitem{byggmastar_effects_2018}
Byggm\"astar J, Granberg F and Nordlund K 2018 {\em Journal of Nuclear
  Materials\/} {\bf 508} 530--539 ISSN 0022-3115

\bibitem{ackland_development_2004}
Ackland G~J, Mendelev M~I, Srolovitz D~J, Han S and Barashev A~V 2004 {\em J.
  Phys.: Condens. Matter\/} {\bf 16} S2629 ISSN 0953-8984

\bibitem{marinica_irradiation-induced_2012}
Marinica M~C, Willaime F and Crocombette J~P 2012 {\em Phys. Rev. Lett.\/} {\bf
  108} 025501

\bibitem{nordlund_molecular_2006}
Nordlund K, Wallenius J and Malerba L 2006 {\em Nuclear Instruments and Methods
  in Physics Research Section B: Beam Interactions with Materials and Atoms\/}
  {\bf 246} 322--332 ISSN 0168-583X

\bibitem{haynes_crc_2015}
Haynes W~M 2015 {\em {{CRC Handbook}} of {{Chemistry}} and {{Physics}}\/} 96th
  ed (Boca Raton, FL: {CRC Press}) ISBN 978-1-4822-6097-7

\bibitem{schepper_positron_1983}
De~Schepper L, Segers D, Dorikens-Vanpraet L, Dorikens M, Knuyt G, Stals L~M
  and Moser P 1983 {\em Phys. Rev. B\/} {\bf 27}(9) 5257--5269
  \urlprefix\url{https://link.aps.org/doi/10.1103/PhysRevB.27.5257}

\bibitem{dezerald_stability_2014}
D\'ezerald L, Marinica M~C, Ventelon L, Rodney D and Willaime F 2014 {\em
  Journal of Nuclear Materials\/} {\bf 449} 219--224 ISSN 0022-3115

\bibitem{klotz_phonon_2000}
Klotz S and Braden M 2000 {\em Phys. Rev. Lett.\/} {\bf 85} 3209--3212 ISSN
  0031-9007, 1079-7114

\bibitem{dever_temperature_1972}
Dever D~J 1972 {\em Journal of Applied Physics\/} {\bf 43} 3293--3301 ISSN
  0021-8979

\bibitem{malerba_comparison_2010}
Malerba L, Marinica M~C, Anento N, Bj\"orkas C, Nguyen H, Domain C, Djurabekova
  F, Olsson P, Nordlund K, Serra A, Terentyev D, Willaime F and Becquart C~S
  2010 {\em Journal of Nuclear Materials\/} {\bf 406} 19--38 ISSN 0022-3115

\bibitem{ma_universality_2019}
Ma P~W and Dudarev S~L 2019 {\em Phys. Rev. Mater.\/} {\bf 3} ISSN 2475-9953

\bibitem{olsson_ab_2016}
Olsson P, Becquart C~S and Domain C 2016 {\em Mater. Res. Lett.\/} {\bf 4}
  219--225 ISSN null

\bibitem{maury_anisotropy_1976}
Maury F, Biget M, Vajda P, Lucasson A and Lucasson P 1976 {\em Phys. Rev. B\/}
  {\bf 14} 5303--5313

\bibitem{ventelon_generalized_2010}
Ventelon L and Willaime F 2010 {\em Philos. Mag.\/} {\bf 90} 1063--1074 ISSN
  1478-6435

\bibitem{gilbert_ab_2010}
Gilbert M~R and Dudarev S~L 2010 {\em Philos. Mag.\/} {\bf 90} 1035--1061 ISSN
  1478-6435, 1478-6443

\end{thebibliography}

\end{document}

\title[Supplementary material]{Supplementary material for: Dynamical stability of radiation-induced C15 clusters in iron}
\author{J Byggmästar$^1$, F Granberg$^1$}
\address{$^1$ Department of Physics, P.O. Box 43, FI-00014 University of Helsinki, Finland}
\ead{jesper.byggmastar@helsinki.fi}

\section{Potential function}

The equations defining the ABOP are listed below. For a more complete description of the parameters involved, see Refs.~\cite{albe_modeling_2002-1,byggmastar_analytical_2019}.
\begin{equation}
V = \sum_i \sum_{j>i} V_{ij} = \sum_i \sum_{j>i} f_{\mathrm{C}}(r_{ij}) [V_{\mathrm{R}}(r_{ij}) - \overline{b}_{ij}V_{\mathrm{A}}(r_{ij})],
\end{equation}

\begin{eqnarray}
V_{\mathrm{R}}(r_{ij}) &= \frac{D_0}{S-1}\exp\left[-\beta\sqrt{2S} (r_{ij}-r_0)\right], \\
V_{\mathrm{A}}(r_{ij}) &= \frac{SD_0}{S-1}\exp\left[-\beta\sqrt{2/S} (r_{ij}-r_0)\right].
\label{eq:abopmorse}
\end{eqnarray}

\begin{equation}
 f_{\mathrm{C}}(r) = \cases{1,& $r \leq R-D$\\
\frac{1}{2} - \frac{1}{2}\sin\left[\frac{\pi}{2D}(r-R) \right],& $|R-r| \leq D$\\
0,& $r \geq R+D$.\\}
\end{equation}

\begin{equation}
\overline{b}_{ij} = \frac{b_{ij}+b_{ji}}{2},
\end{equation}

\begin{equation}
b_{ij} = (1 + \chi_{ij})^{-1/2}.
\end{equation}

\begin{equation}
\chi_{ij} = \sum_{k (\neq i, j)} f_{\mathrm{C}}(r_{ik}) g_{ik}(\theta_{ijk}) \omega_{ijk} \exp\left[\alpha_{ijk}(r_{ij}-r_{ik})\right],
\label{eq:chi}
\end{equation}

\begin{equation}
g_{ik}(\theta_{ijk}) = \gamma_{ik} \left[1 + \frac{c_{ik}^2}{d_{ik}^2} - \frac{c_{ik}^2}{d_{ik}^2 + (h_{ik}+\cos\theta_{ijk})^2}\right].
\end{equation}

The ABOP is joined with the the universal repulsive Ziegler-Biersack-Littmark potential~\cite{ziegler_stopping_1985}, $V_{\mathrm{ZBL}}(r_{ij})$, according to
\begin{equation}
V'_{ij} = F(r_{ij}) V_{ij} + \left[1 - F(r_{ij})\right] V_{\mathrm{ZBL}}.
\label{eq:reppot}
\end{equation}
where
\begin{equation}
F(r) = \frac{1}{1+\exp[-b_{\mathrm{f}}(r-r_{\mathrm{f}})]}.
\label{eq:reppot2}
\end{equation}

\section{Notes on the potential refitting}

\begin{table}
 \centering
 \begin{tabular}{ll}
  \toprule
  Parameter & Value \\
  \midrule
  $D_0$ (eV) & \textbf{1.465} \\
  $r_0$ (Å) & \textbf{2.3} \\
  $\beta$ & 1.4 \\
  $S$ & 2.0693109 \\
  $\gamma$ & 0.0115751 \\
  $c$ & 1.2898716 \\
  $d$ & 0.3413219 \\
  $h$ & \textbf{-0.38} \\
  $R$ (Å) & \textbf{3.26} \\
  $D$ (Å) & 0.2 \\
  $\alpha$ & \textbf{0.8} \\
  $\omega$ & 1.0 \\
  $b_{\mathrm{f}}$ & \textbf{10.0} \\
  $r_{\mathrm{f}}$ & \textbf{1.2} \\
  \bottomrule
 \end{tabular}
 \caption{Parameters of the revised Fe--Fe ABOP. Values in bold have been modified, others are identical to the earlier parameter set from Refs.~\cite{muller_analytic_2007,bjorkas_comparative_2007}}
 \label{tab:potpars}
\end{table}

As mentioned in the main article, our starting point when developing the potential was the existing ABOP parametrisation by M\"uller et al.~\cite{muller_analytic_2007} with the short-range addition by Björkas and Nordlund~\cite{bjorkas_comparative_2007} (which we will call ''ABOP-MEABN'', and the revised version simply ''ABOP''). The goal was to adjust the parameters in order to improve the energetics of single interstitials and vacancies, including migration energies, and to (more importantly) correctly stabilise small C15 clusters relative to dislocation loops in accordance with DFT predictions. To this end, the parameters affecting these properties were iteratively and manually adjusted until a satisfactory agreement with the target data was achieved. The stability of C15 clusters was achieved by changing the optimal interatomic angle defined by the parameter $h$, which was adjusted to reproduce the cohesive energy of bulk C15 and formation energies of small di- and tetra-SIA C15 clusters in bcc iron. Reproducing the relative stability of C15 clusters and dislocation loops in quantitative agreement with DFT for all sizes was not possible. Decreasing $h$ further can increase the stability of large C15 clusters towards DFT accuracy, but consequently leads to poor elastic constants of bulk bcc and too-stable smaller C15 clusters.

Furthermore, we found that the energies of single and small SIA configurations as well as the vacancy formation energy could be improved by increasing the value of the $\alpha$ parameter, which initially had been set to zero by M\"uller et al. A non-zero $\alpha$ enables a dependence on the relative bond lengths in the three-body contribution. M\"uller et al. noted that using a non-zero $\alpha$ resulted in a reduced thermal stability of the fcc phase. However, since we are here mainly interested in radiation damage in bcc iron, sacrificing the stability of the fcc phase of iron is acceptable in favour of significantly improving the energetics of small defect configurations.

\begin{figure}
 \centering
 \includegraphics[width=\linewidth]{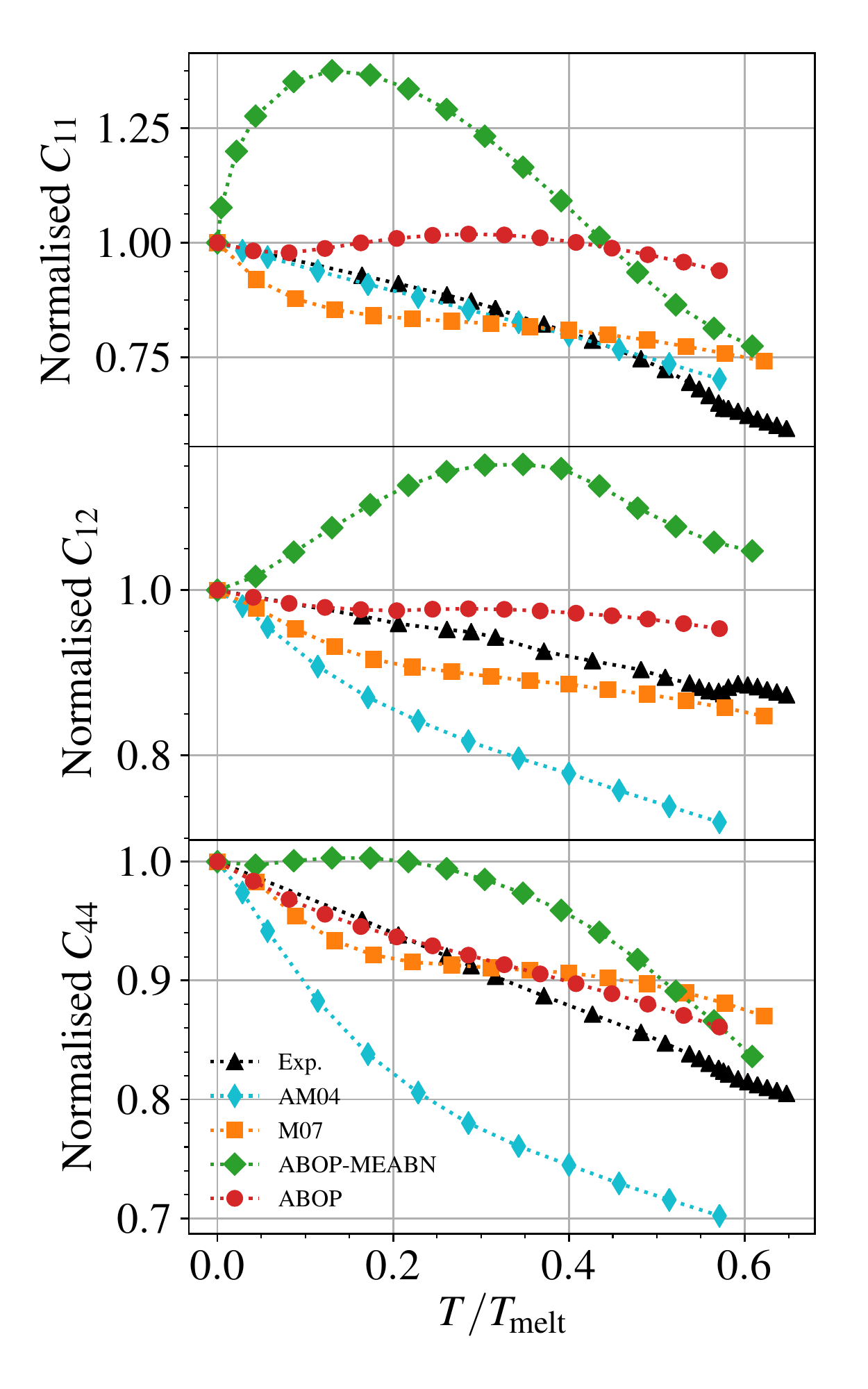}
 \caption{Temperature dependence of elastic constants compared with experimental data from Ref.~\cite{dever_temperature_1972}. Since the potentials were fitted to slightly different experimental data, the elastic constants are normalised by the value at 0 K. The temperatures are also given as fractions of the melting temperature, due to different melting points in the potentials.}
 \label{fig:elastic}
\end{figure}

The cutoff range of the original parametrisation by M\"uller, and especially following the short-range modification in~\cite{bjorkas_comparative_2007}, is very close to the second-nearest neighbour in the bcc crystal, leading to peculiar temperature effects (see Fig.~\ref{fig:elastic}). We therefore extended the cutoff range slightly in order to achieve a reasonable thermal expansion and better elastic properties at finite temperatures. The new cutoff radius was chosen to still be between the second and third nearest neighbour atom in bcc, while also being in-between neighbour shells in e.g. fcc and bulk C15 to avoid unphysical effects. In addition, the cutoff range was carefully chosen to remove spurious energy barriers in the migration paths of vacancies and self-interstitials. Sudden changes in the total coordination of the migrating atom due to atoms entering and exiting the cutoff range were seen to lead to unphysical peaks in the migration barriers.

Finally, to account for the above parameter adjustments, we slightly modified the values of the dimer energy and bond length ($D_0$ and $r_0$) to reproduce good cohesive energies and lattice constants of bcc iron and other phases. All unmentioned parameters were kept unchanged, including the parameters of the angular function (except the parameter $h$), and the parameters controlling the stiffness and the dependence on coordination ($\beta$, $S$, and $\gamma$). We note that the adjustments we have made eliminates the spontaneous bcc--fcc phase transition, which is correctly reproduced by the original parametrisation.

\begin{figure}
 \centering
 \includegraphics[width=\linewidth]{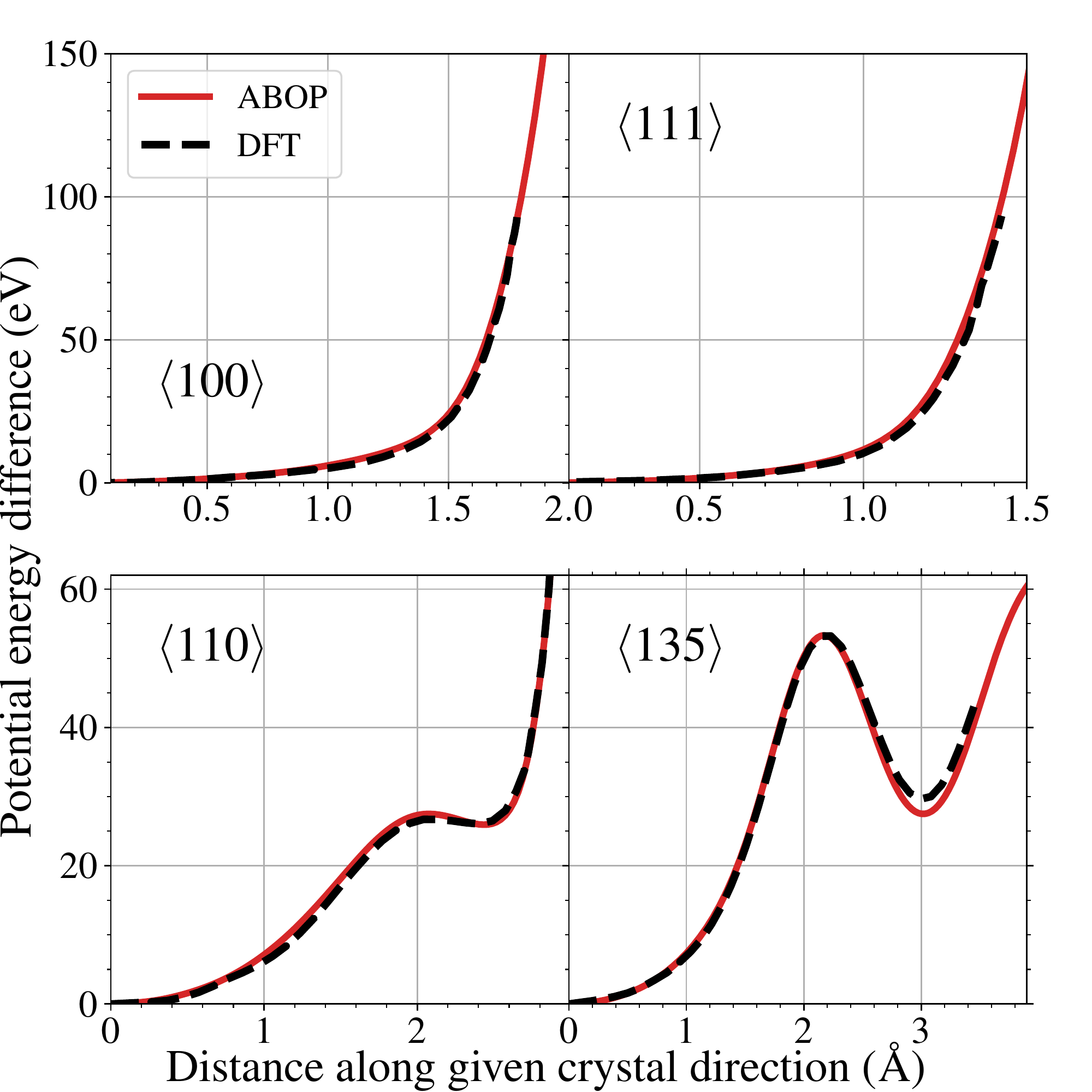}
 \caption{Energy difference for step-wise static movement of one atom along different crystal directions in a rigid bcc lattice. DFT data are from Ref.~\cite{olsson_ab_2016}.}
 \label{fig:qsd}
\end{figure}

The short-range part of the original Fe ABOP introduced in~\cite{bjorkas_comparative_2007} was also changed by modifying the parameters of the Fermi function used for the smooth transition from the ZBL potential to the Tersoff-like near-equilibrium part. While the previous Fermi parameters lead to reasonable threshold displacement energies, the many-body interaction curves when moving an atom in a rigid lattice along a given crystal direction (also called ''quasi-static drag'' or ''sudden relaxation'' curves) are not in agreement with DFT results. The new parameters of the Fermi function were chosen so that these repulsive many-body interaction curves are accurately reproduced (Fig.~\ref{fig:qsd}). Previously, when modifying the short-range part of an EAM potential~\cite{byggmastar_effects_2018}, we used these many-body interaction curves to validate the potential after having obtained reasonable threshold displacement energies. Here, we found that a reverse approach is much more efficient. That is, tuning the short-range potential to reproduce the repulsive many-body energies from DFT automatically resulted in good threshold displacement energies (as long as other relevant properties, such as the formation energy and volume of the lowest-energy self-interstitial configuration, are well reproduced).

The original ABOP overestimates the melting temperature by about 500 K. Our revised version produces an even higher melting point, about 2450 K compared to the experimental value 1811 K. The significant overestimation of the melting point may affect the results of collision cascades by limiting the mixing efficiency during the heat spike. However, lowering the melting point is not possible without affecting more important properties.

\section{Benchmarking the revised interatomic potential}
\label{sec:results}

In what follows, we show the results obtained from benchmarking tests of the revised ABOP. Most results are compared with the previous Fe ABOP (ABOP-MEABN) and two EAM potentials (AM04~\cite{ackland_development_2004} and M07-B~\cite{marinica_irradiation-induced_2012,byggmastar_effects_2018}).

The formation and binding energies of single and small clusters of self-interstitial atoms (SIAs) and vacancies were calculated by relaxing non-cubic bcc systems containing the defects and around 2\,000 atoms. All defect configurations were relaxed to a force convergence of $10^{-8}$ eV/Å at zero pressure.

Threshold displacement energies (TDEs) were simulated using the same methods as described in details in~\cite{nordlund_molecular_2006,byggmastar_effects_2018}. A total of 5\,000 random directions were sampled to obtain full angular map of the TDEs.

\begin{table}[h]
 \centering
 \caption{Bulk properties of bcc iron compared with experimental data.}
 \label{tab:bcc}
 \begin{threeparttable}
  \begin{tabular}{lll}
   \toprule
   & Exp. & ABOP \\
   \midrule
    $E_\mathrm{coh}$ & 4.31\tnote{a} & 4.34867  \\
    $a$ & 2.866\tnote{a} & 2.85988 \\
    $B$ & 169\tnote{a} & 171 \\
    $c_{11}$ & 226\tnote{a} & 223 \\
    $c_{12}$ & 140\tnote{a} & 145 \\
    $c_{44}$ & 116\tnote{a} & 118 \\
%    $\gamma_{1jk}$ & 147 & 105, 84, 117 \\
    $T_\mathrm{melt}$ & 1811\tnote{a} & $2450 \pm 50$ \\
    $E_\mathrm{vac}^\mathrm{f}$ & 1.4--2.0\tnote{b} & 1.70 \\
   \bottomrule
  \end{tabular}
  \begin{tablenotes}
   \item[a] Ref.~\cite{haynes_crc_2015}
   \item[b] Ref.~\cite{schepper_positron_1983} 
  \end{tablenotes}
 \end{threeparttable}
\end{table}

\begin{table*}
 \centering
 \caption{Properties of the bulk phases closest in energy to the bcc ground state. $\Delta E$ (eV) is the difference in cohesive energy per atom compared to the bcc phase, $a$ is the lattice constant (Å), and $B$ is the bulk modulus (GPa).}
 \label{tab:bulk}
  \begin{tabular}{llllll}
   \toprule
   & DFT~\cite{muller_analytic_2007,dezerald_stability_2014} & ABOP & ABOP-MEABN & M07-B & AM04 \\
   \midrule
    \textbf{fcc} \\
   %\midrule
    $\Delta E$ & 0.11 & 0.088 & 0.028 & 0.12 & 0.12 \\
    $a$ & 3.52 & 3.61 & 3.65 & 3.70 & 3.66 \\
    $B$ &  & 165 & 163 & 68 & 65 \\
%    $E_\mathrm{Bain}$ (meV/atom) & 45 & 8 & 41 & 0.0 & 0.6 \\
    \midrule
    \textbf{hcp} \\
    $\Delta E$ &  0.06 & 0.086 & 0.024 & 0.096 & 0.116 \\
    $a$ & 2.48 & 2.56 & 2.58 & 2.62 & 2.61 \\
    $c$ & 3.93 & 4.17 & 4.20 & 4.31 & 4.17 \\
    $B$ &  & 165 & 163 & - & 73 \\
   \midrule
    \textbf{C15} \\
    $\Delta E$ & 0.15 & 0.14 & 0.20 & 0.22 & 0.18 \\
    $a$ & 6.64 & 6.64 & 6.73 & 6.51 & 6.66 \\
    $B$ & 150 & 157 & 154 & 121 & 107 \\
   \midrule
    \textbf{A15} \\
    $\Delta E$ & 0.09 & 0.057 & 0.11 & 0.079 & 0.085 \\
    $a$ & 4.56 & 4.56 & 4.62 & 4.45 & 4.53 \\
    $B$ &  & 166 & 159 & 212 & 126 \\
%   \midrule
%    \textbf{sc} \\
%    $\Delta E$ &  & 0.75 & 1.18 & 1.15 & 0.68 & 0.70 \\
%    $a$ &  & 2.39 & 2.24 & 2.45 & 2.25 & 2.24 \\
%    $B$ &  &  & 444 & 89 & 407 & 250 \\
%   \midrule
%    \textbf{DIA} \\
%    $\Delta E$ &  & 1.17 & 2.29 & 2.09 & 1.78 & 1.42 \\
%    $a$ &  & 4.856 & 5.60 & 5.64 & 5.43 & 5.45 \\
%    $B$ &  &  & 39 & 40 & 88 & 43 \\
   \bottomrule
  \end{tabular}
\end{table*}

%\subsection{Vibrational and thermal properties}

\begin{figure}[h]
 \centering
 \includegraphics[width=\linewidth]{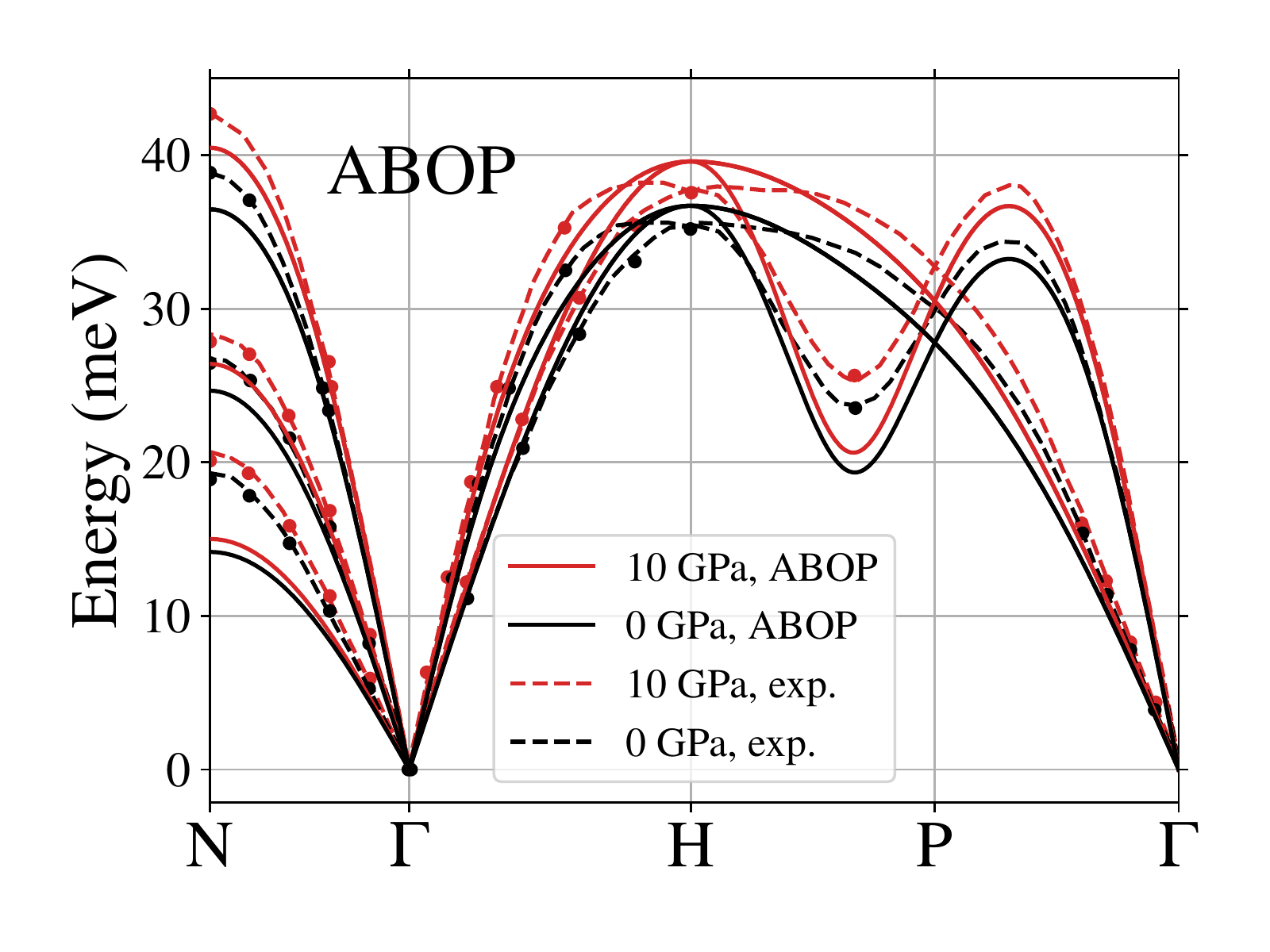}
 \includegraphics[width=\linewidth]{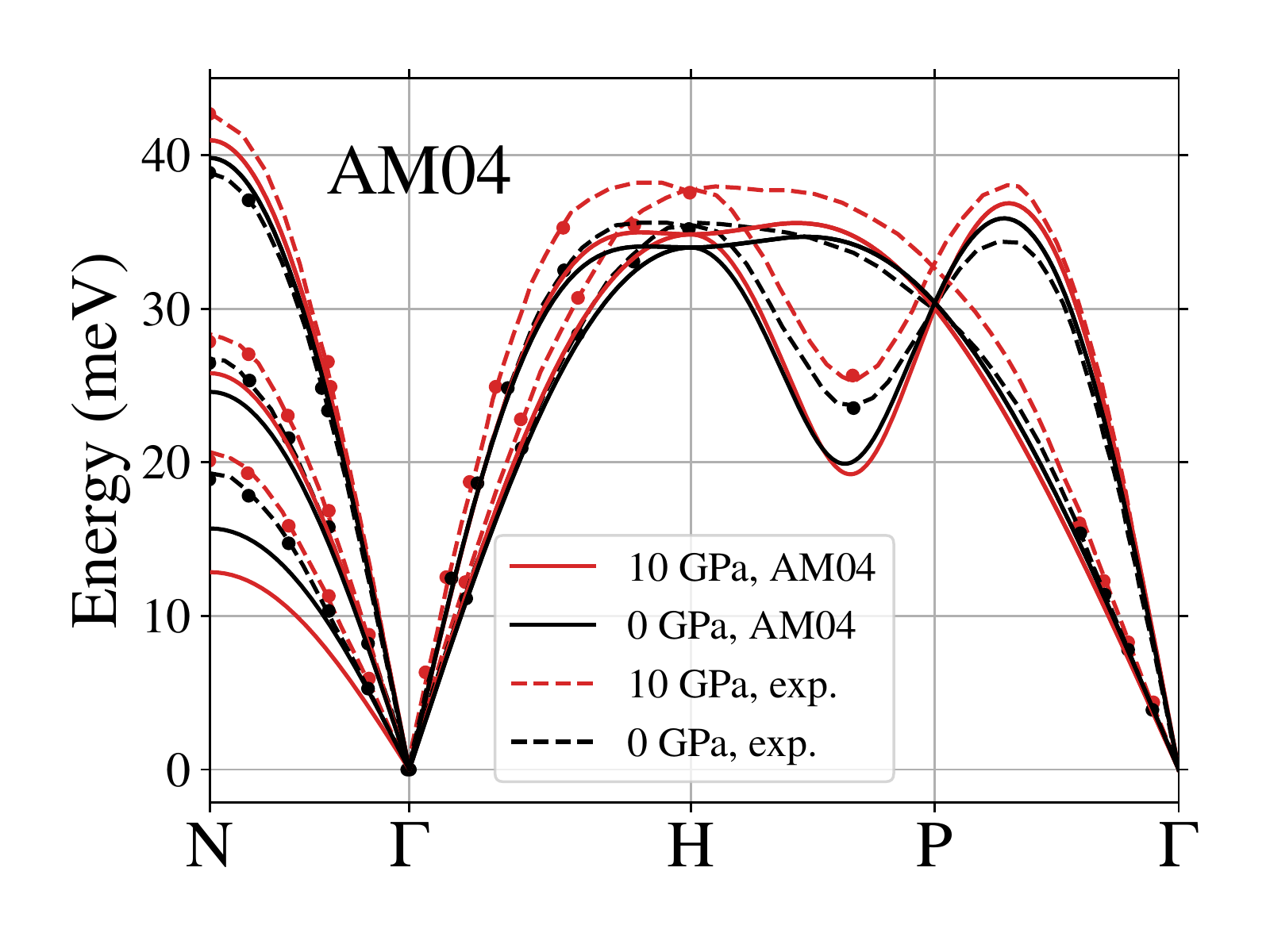}
 \caption{Phonon dispersions of bcc iron at 0 and 10 GPa compared with experimental data from Ref.~\cite{klotz_phonon_2000}. The new ABOP predicts a pressure dependence similar to the experimental observations, i.e. a 5--10\% stiffening of all branches. Most EAM potentials show a softening of some branches and stiffening of others (see e.g the lowest acoustic branch at the N point).}
 \label{fig:phonon}
\end{figure}

%\subsection{Defect energetics}

\begin{table*}
 \centering
 \caption{Self-interstitial atom (SIA) formation energies (eV), relaxation volumes (in units of the relaxed atomic volume $\Omega_0$), and migration energies of a single vacancy and a \hkl<110> SIA (eV). Stars indicate relaxation to a lower-energy site unless constrained.}
 \label{tab:single_def}
  \begin{tabular}{llllll}
   \toprule
   & DFT~\cite{malerba_comparison_2010,ma_universality_2019} & ABOP & ABOP-MEABN & M07-B & AM04 \\
   \midrule
   $E_\mathrm{f}^{\langle 110 \rangle}$ & 3.77, 3.94, 4.32 & 3.99 & 4.52 & 3.69 & 3.52  \\
   $E_\mathrm{f}^{\langle 111 \rangle}$ & 4.49, 4.66, 5.09 & 4.47* & 4.78 & 4.36* & 4.01* \\
   $E_\mathrm{f}^{\langle 111 \rangle \mathrm{crowdion}} - E_\mathrm{f}^{\langle 111 \rangle \mathrm{db}}$ & 0.02, 0.001, 0.003 & 0.035 & $-0.12$ & 0.002 & 0.007 \\
   $E_\mathrm{f}^{\langle 100 \rangle}$ & 4.80, 5.04, 5.46 & 5.63* & 5.77 & 4.76* & 4.35* \\
   $E_\mathrm{f}^\mathrm{octa}$ & 4.97, 5.68, 5.56 & 5.30 & * & 4.90* & 4.17 \\
   $E_\mathrm{f}^\mathrm{tetra}$ & 4.28, 4.88, 4.79 & 4.94* & 5.28 & 4.31* & 4.15* \\
   \midrule
   $\Omega_\mathrm{rel}^{\hkl<110>}$ & 1.62 & 1.63 & 1.22 & 1.48 & 1.24 \\
   $\Omega_\mathrm{rel}^\mathrm{vac}$ & $-0.22$ & $-0.35$ & $-0.36$ & $-0.10$ & $-0.22$ \\
   \midrule
   $E_\mathrm{m}^{\mathrm{vac}}$ & 0.67 & 0.70 & 0.92 & 0.69 & 0.64  \\
   $E_\mathrm{m}^{\hkl<110>}$ & 0.34 & 0.345 & 0.15 & 0.29 & 0.31 \\
   \bottomrule
  \end{tabular}
\end{table*}

\begin{table*}
 \centering
 \caption{Formation energies of different configurations of multiple SIAs. Formation energies are given for the parallel \hkl<110> configurations. For all other configurations, the difference in energy compared to \hkl<110>$_x$ are given. Stars indicate relaxation to a lower-energy site unless constrained.}
 %DFT refs: Dezerald for C15, Malerba2010 (see refs, Terentyev) most. penta refs from different DFT (lower than other values). 100 always rotates to 110. tetra 110 and 111 slightly off-axis and non-parallel in ABOP-new. tetra-111 in M07 also slightly off-axis non-parallel (E difference therefore somewhat justified, to be fair). Most tetra and penta slightly tilted, non-parallel.
 \label{tab:multi_SIA}
  \begin{tabular}{llllll}
   \toprule
   & DFT~\cite{malerba_comparison_2010,dezerald_stability_2014} & ABOP & ABOP-MEABN & M07-B & AM04 \\
   \midrule
    $\hkl<110>_2$ & 6.99--7.55 & 6.99 & 8.20 & 6.30 & 6.21  \\
    $\hkl<110>_2$ ring & $-0.1$ & 0.20 & $-0.26$ & 0.12 & 0.25  \\
    $\hkl<111>_2$ & 0.75 & 0.53 & 0.00 & 0.99 & 0.62  \\
    $\hkl<100>_2$ & 0.11 & * & 1.69* & * & *  \\
    C15$_2$ & 0.8 & 0.42 & 0.93 & 0.28 & 1.37  \\
   \midrule
    $\hkl<110>_3$ & 9.89--10.39 & 9.90 & 11.51 & 8.94 & 8.83 \\
    $\hkl<110>_3$ ring & $-0.06$ & 0.65 & 0.10 & 0.27 & 0.83 \\
    $\hkl<111>_3$ & 0.36 & 0.35 & $-0.23$ & * & 0.67 \\
    $\hkl<100>_3$ & 1.62 & * & * & * & * \\
   \midrule
    $\hkl<110>_4$ & 12.31--13.60 & 12.18 & 15.08 & 10.93 & 10.97 \\
    $\hkl<111>_4$ & 0.11 & 0.06 & $-1.63$ & 0.78 & 0.17 \\
    $\hkl<100>_4$ & 1.07 & * & * & * & 1.06 \\
    C15$_4$ & $-1.29$, $-1.83$ & $-1.62$ & $-1.73$ & $-1.98$ & $-0.24$ \\
   \midrule
    $\hkl<110>_5$ & 14.18 & 14.91 & 16.92 & 13.40 & 13.39 \\
    $\hkl<111>_5$ & $-0.3$ & $-0.12$ & $-0.78$ & 0.43 & $-0.01$ \\
    $\hkl<100>_5$ & 2.27 & * & * & * & * \\
   \bottomrule
  \end{tabular}
\end{table*}

\begin{table*}
 \centering
 \caption{Binding energies of small vacancy clusters (only the most stable configuration is given).}
 \label{tab:Eb_vac}
  \begin{tabular}{llllll}
   \toprule
   & DFT~\cite{malerba_comparison_2010} & ABOP- & ABOP-MEABN & M07-B & AM04 \\
   \midrule
   $V$--$V$ & 0.23--0.30 & 0.23 & 0.20 & 0.33 & 0.24 \\
   $V_2$--$V$ & 0.36--0.37 & 0.49 & 0.32 & 0.33 & 0.30 \\
   $V_3$--$V$ & 0.62--0.70 & 0.73 & 0.60 & 0.75 & 0.58 \\
   \bottomrule
  \end{tabular}
\end{table*}

\begin{figure}
 \centering
 \includegraphics[width=\linewidth]{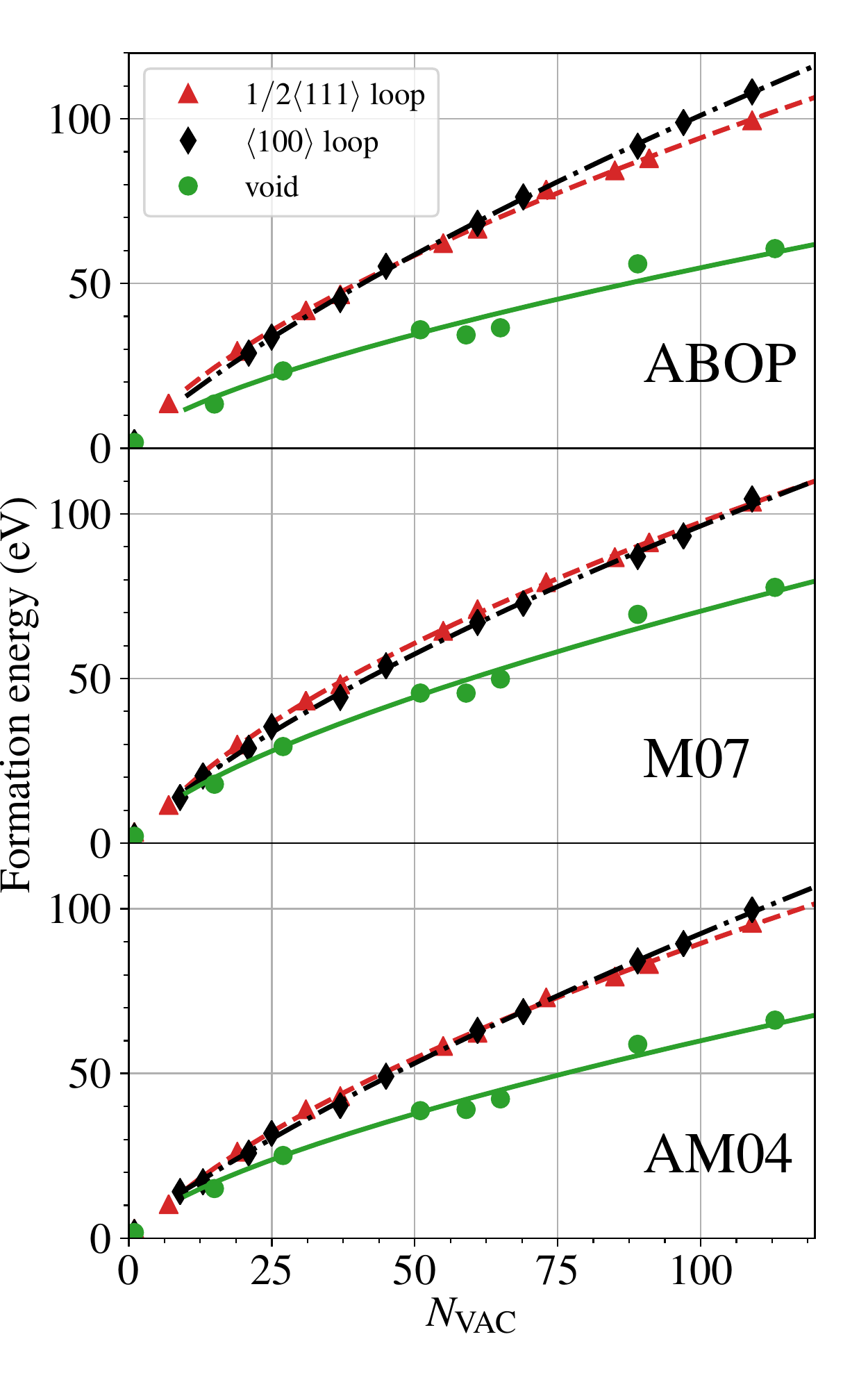}
 \caption{Formation energies of vacancy-type clusters.}
 \label{fig:Ef_vac}
\end{figure}

%\subsection{Short-range properties}

\begin{figure}
 \centering
 \includegraphics[width=\linewidth]{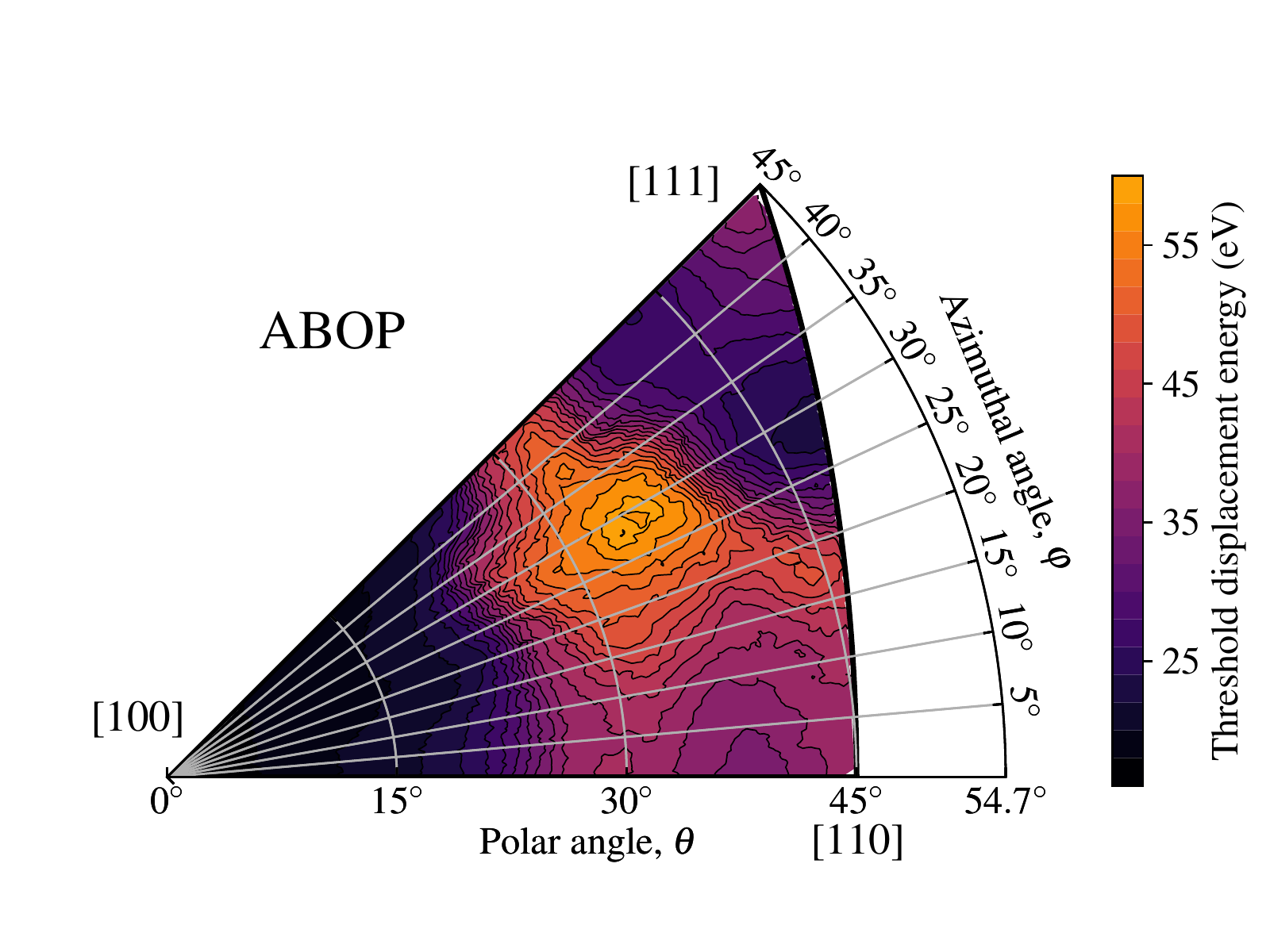}
 \includegraphics[width=\linewidth]{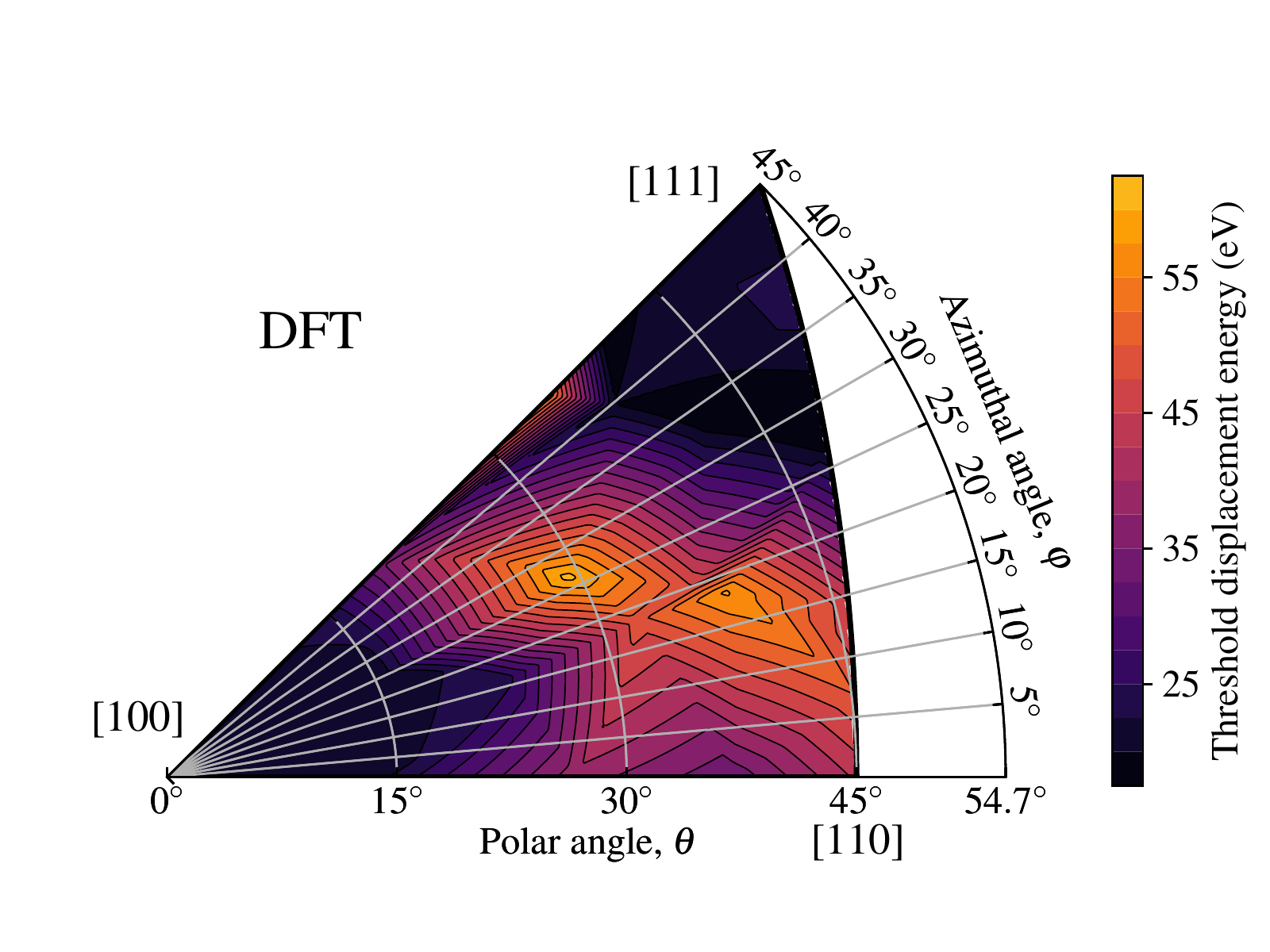}
 \caption{Angular maps of threshold displacement energies in the ABOP (at and compared with DFT results from Ref.~\cite{olsson_ab_2016}. The global average in the ABOP is $35.2 \pm 0.2$ eV, compared to 32 eV in DFT.}
 \label{fig:TDEmaps}
\end{figure}

\begin{figure}
 \centering
 \includegraphics[width=\linewidth]{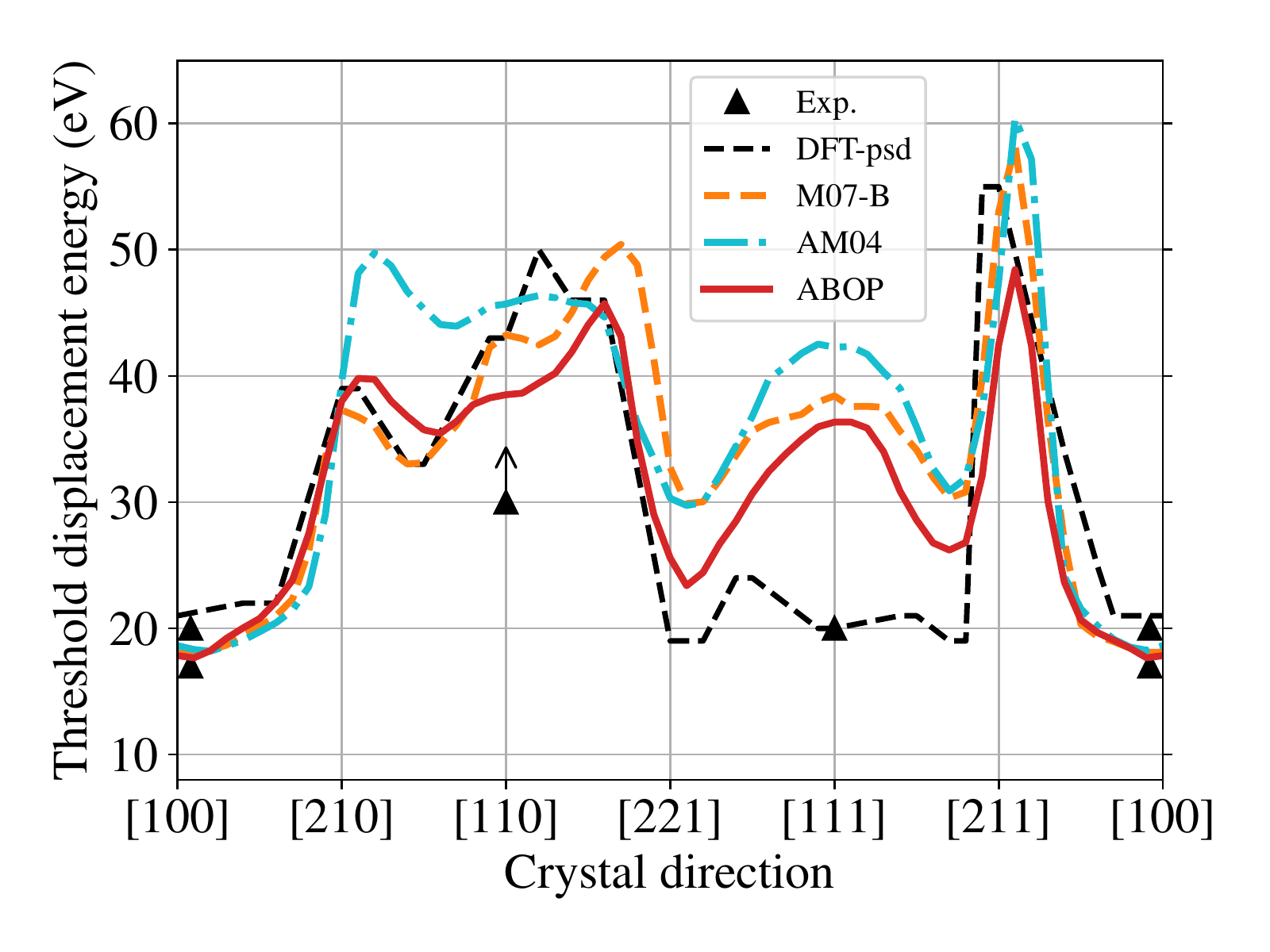}
 \caption{Threshold displacement energies. The DFT-psd data are from Ref.~\cite{olsson_ab_2016}. The arrow at the experimental value at \hkl[110] indicate the estimated value $>30$ eV~\cite{maury_anisotropy_1976}.}
 \label{fig:TDE-1D}
\end{figure}

%\subsection{Extended defects}

\begin{figure}
 \centering
 \includegraphics[width=\linewidth]{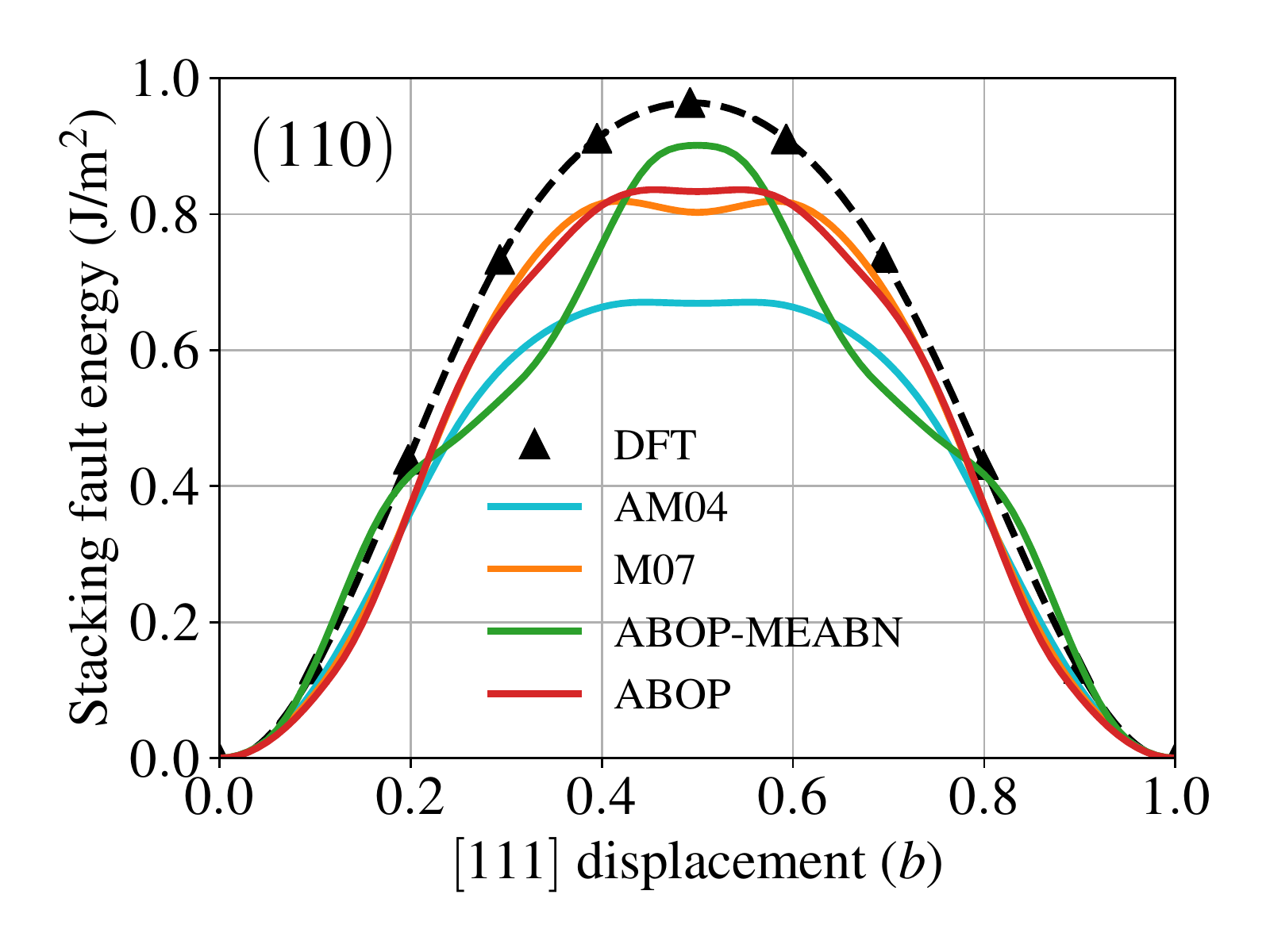}
 \includegraphics[width=\linewidth]{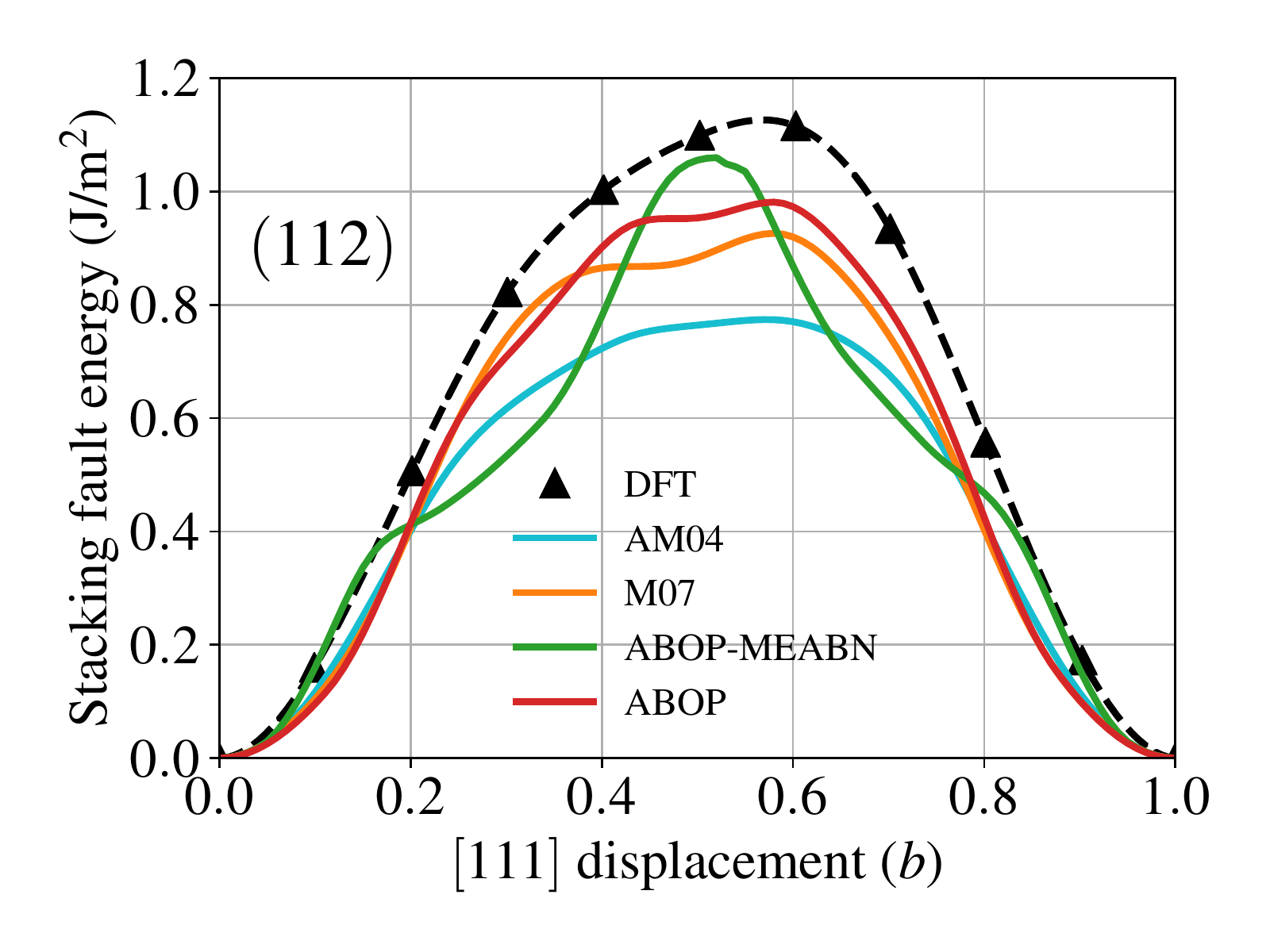}
 \caption{Generalised stacking fault energies. DFT data are from Ref.~\cite{ventelon_generalized_2010}.}
 \label{fig:stackfault}
\end{figure}

\begin{figure}
 \centering
 \includegraphics[width=0.85\linewidth]{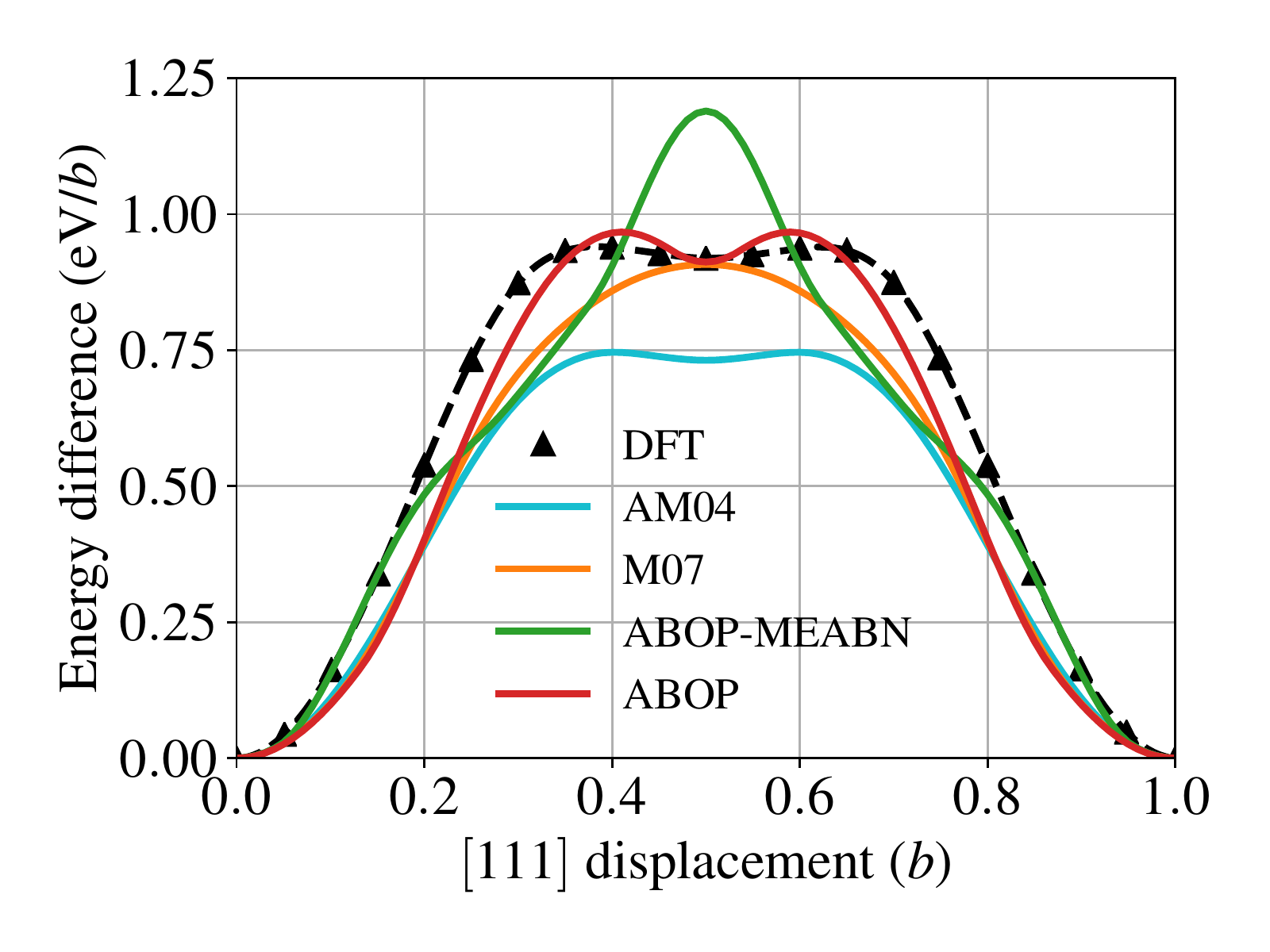}
 \caption{$[111]$ string displacement. DFT data are from Ref.~\cite{gilbert_ab_2010}.}
 \label{fig:111string}
\end{figure}

\begin{figure}
 \centering
 \includegraphics[width=0.8\linewidth]{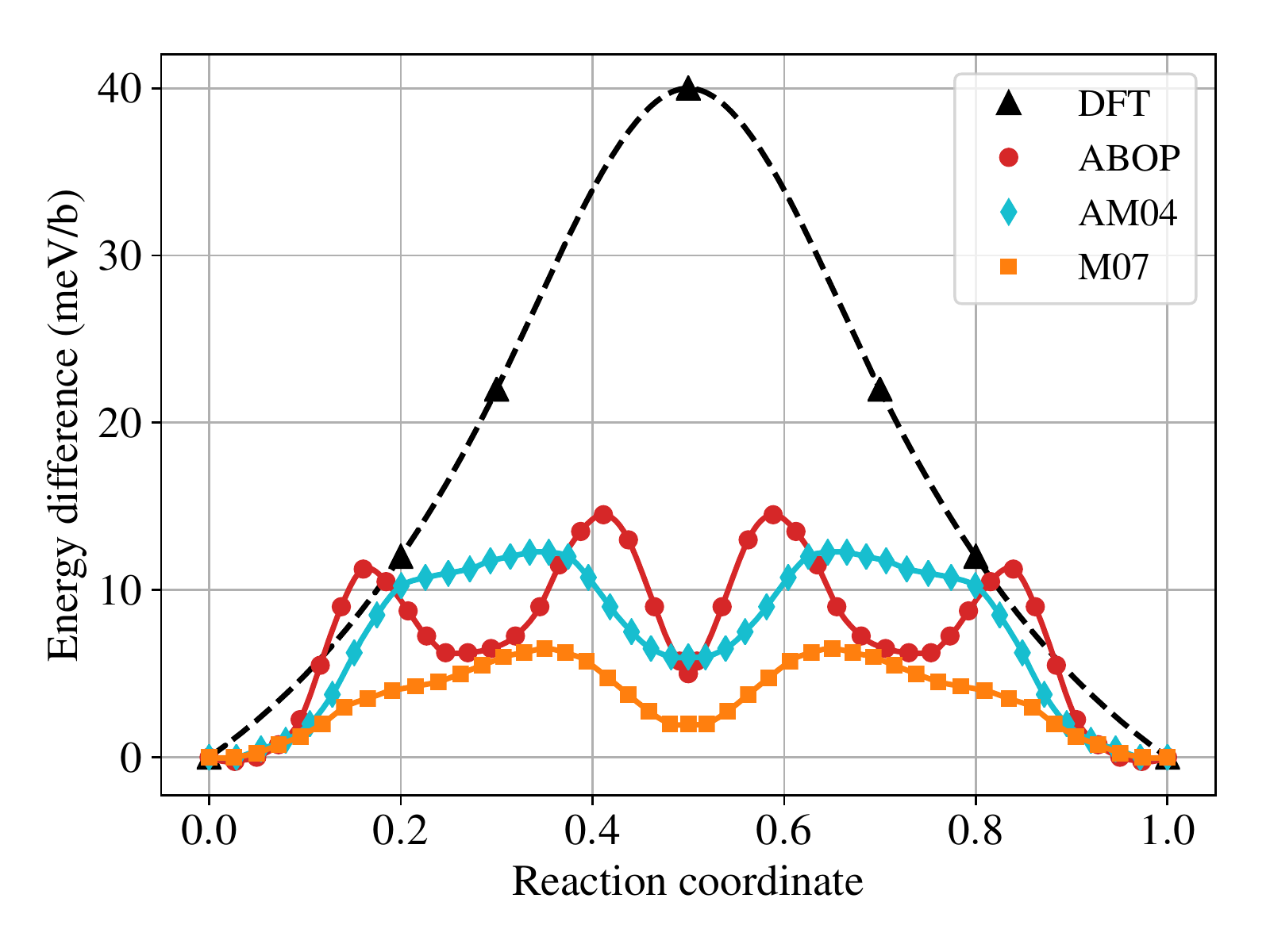}
 \caption{Peierls barrier of screw dislocation movement. Interestingly, despite reproducing the shape and saddle point energy of the \hkl[111] string displacement (Fig.~\ref{fig:111string}) and the stacking fault energy profiles (Fig.~\ref{fig:stackfault}), the ABOP fails completely in reproducing the corresponding shape and saddle point energy of the Peierls barrier. While all potentials shown here are unable to reproduce the DFT profile, they all, however, predict the correct compact non-degenerate core structure.}
 \label{fig:peierls}
\end{figure}

\clearpage
\section*{References}
\bibliography{mybib}